\documentclass[aps, prb, superscriptaddress, reprint]{revtex4-2}
\usepackage{graphicx}
\usepackage{dcolumn}
\usepackage{bm}
\usepackage{amsfonts}
\usepackage{amsmath}
\usepackage{amssymb}

\usepackage{color}

\usepackage{graphicx,hyperref}
\hypersetup{%
  breaklinks = {true},
  citecolor = {blue},
  colorlinks = {true},
  linkcolor = {red},
}

\def \hc{\hat{c}}
\def \hrho{\hat{\rho}}

\begin{document}


\title{Anomalously large relaxation times in dissipative lattice models beyond the non-Hermitian skin effect}

\author{Gideon Lee }
\affiliation{Pritzker School of Molecular Engineering, The University of Chicago, Chicago, Illinois 60637, USA}

\author{Alexander McDonald}
\affiliation{Pritzker School of Molecular Engineering, The University of Chicago, Chicago, Illinois 60637, USA}
\affiliation{Institut Quantique \& Département de Physique, Université de Sherbrooke, Sherbrooke, Québec, J1K 2R1, Canada}
\affiliation{Department of Physics, University of Chicago, Chicago, IL 60637, USA}

\author{Aashish Clerk}
\affiliation{Pritzker School of Molecular Engineering, The University of Chicago, Chicago, Illinois 60637, USA}

\date{\today}

\begin{abstract}
We show for generic quantum non-Hermitian tight-binding models that relaxation timescales of local observables are not controlled by the localization length $\xi_{\rm loc}$ associated with the non-Hermitian skin effect, contrary to popular belief. Instead, interference between eigenvectors effectively makes the extreme localization of modes largely irrelevant to relaxation; this is ultimately a consequence of causality and locality. Focusing on the paradigmatic Hatano-Nelson model, we demonstrate that there exists instead a much larger length scale $\xi_{\rm prop}$ which controls the rate of decay towards the steady state. Further, varying $\xi_{\rm prop}$ can lead to anomalously large relaxation times that scale with system size, or to the expected behavior where the dissipative gap correctly predicts the rate of decay. Our work highlights an important aspect of the non-Hermitian skin effect: the exceptional sensitivity to boundary conditions here necessarily takes a finite amount of time to manifest itself.   
\end{abstract}

\maketitle

\section{Introduction}

The exotic effects of non-Hermiticity in classical and quantum systems alike has generated widespread interest in recent years \cite{Okuma_2022_NH_review, Ashida_2020_NH_review, Bergholtz_2021_RMP_NH_review}. One such unusual phenomenon is the existence of anomalously long relaxation times: generic local observables reach their steady-state value much more slowly than the characteristic decay rates, the smallest of which is known as the dissipative gap, would suggest \cite{wolf_math_phys_paper, Song_2019_PRL_chiral_damping}. This behavior has been observed in a wide range of  models such as random quantum circuits \cite{Bensa_PRX_2021_randcircs, Znidaric_2022_arxiv_randcirc_NHSE, Bensa_PRR_2022_two_step_phantom, Claeys_2022_randcirc} and non-Hermitian tight-binding models \cite{Haga_PRL_2021_LSE, Mori_PRL_2020_LSE, Yang_2022_PRR}. The latter 
also exhibits counter-intuitive behavior known as the non-Hermitian skin effect (NHSE) \cite{NHSE_ref_1, NHSE_ref_2, NHSE_ref_3, Kunst_2018_PRL_biorth_correspondence}, which occurs when the Hamiltonian under open boundary conditions has a macroscopic number of localized edge modes and a drastically different spectrum than its periodic boundary condition counterpart. The NHSE has been probed in experiments \cite{Coulais2019, liang2022_nhexpt, wang2022_nhexpt}. 

Recent work has suggested that these two counterintuitive effects are in fact intimately related. By interpreting a Lindblad superoperator as a non-Hermitian Hamiltonian, Ref.~\cite{Haga_PRL_2021_LSE} argues that exponentially-large-in-system-size expansion coefficients due the NHSE leads to a unexpectedly-large relaxation time 
\begin{align}\label{eq:naive_relaxation_intro}
\tau_{\rm EVec}
{\sim}
 \frac{L}{\xi_{\rm loc} \Delta},
\end{align}
where $\tau$ is a reasonably-defined relaxation timescale for a local observable, $L$ is the size of the system, $\Delta$ is the dissipative gap \cite{Minganti2018_PRA_Liouvillian_diss_gap}, and $\xi_{\rm loc}$ is the localization length of the corresponding least-damped Liouvillian eigenmode. Since the argument leading to Eq.~(\ref{eq:naive_relaxation_intro}) is based solely on the localized Lindblad eigenvectors, we refer to it as the eigenvector (EVec) prediction. A similar argument was put forth in Refs.~\cite{Mori_PRL_2020_LSE, Mori_PRR_2021_MB_explosion}. 

In this work, we argue that this analysis is not valid for generic 1D non-Hermitian tight-binding models; 
not only is the relaxation time insensitive to $\xi_{\rm loc}$, but $\tau \Delta$ need not scale with $L$ even when the model exhibits the NHSE. The mechanism leading to the breakdown of Eq.~(\ref{eq:naive_relaxation_intro}) is simple but ubiquitous in tight-binding models exhibiting the NHSE: the dynamics of local observables involve a large number of eigenvectors, and interference between these modes can effectively make their exponentially-localized nature largely irrelevant. This conclusion, unlike the prediction of Eq.~(\ref{eq:naive_relaxation_intro}), cannot be reached by considering the eigenvectors or eigenvalues separately. Rather, the combination of both is essential to understand the physics at play, as is done naturally when one considers the system's Green's functions.  While we focus on a few well-studied models (the quantum realization of the Hatano-Nelson model \cite{OG_QHN_1, OG_QHN_2} and two additional models in App.~\ref{app:other_models} for concreteness, we argue that this cancellation is generic and must occur due to locality. 
Note that one can also study anomalous relaxation in non-Hermitian models using pseudo-spectral methods \cite{Flynn_2021_majorana_bosons, Okuma_Sato_PRB_2020_pseudospectra, Znidaric_2022_arxiv_randcirc_NHSE}; however, these do not generally allow one to make precise statements, nor do they provide the intuition we present here.

\begin{figure*}[htpb]
    \centering
    \includegraphics[width=\linewidth]{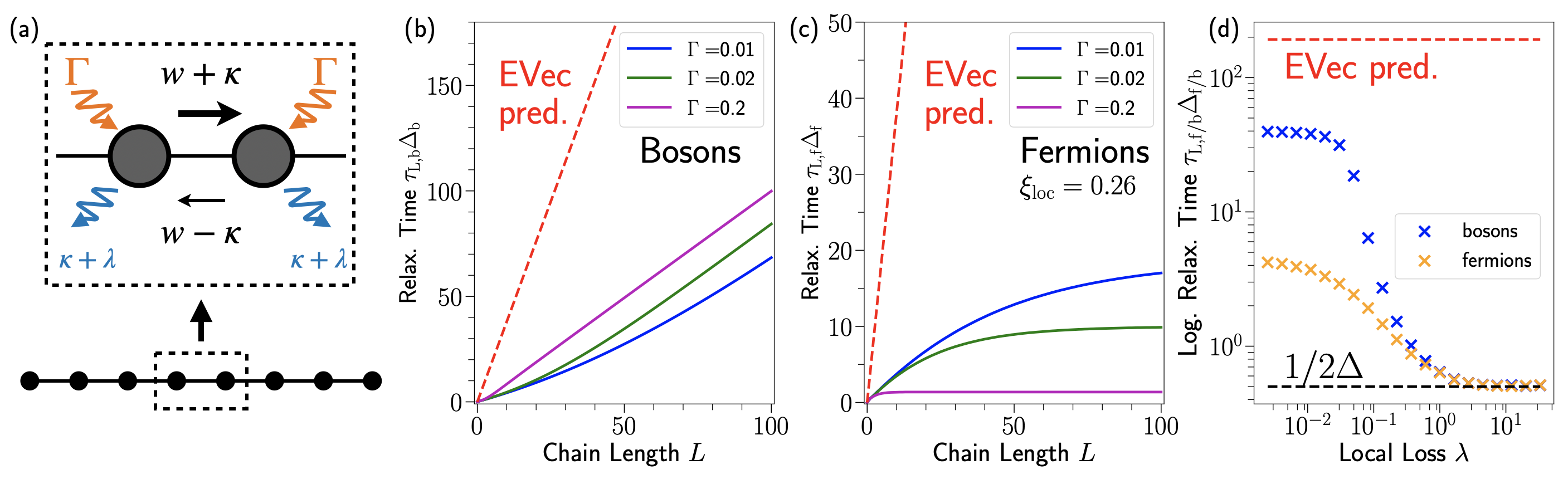}
    \caption{(a) Schematic of the quantum Hatano-Nelson (QHN) model realized using engineered dissipation with hopping amplitudes $w\pm \kappa$, local pumping $\Gamma$ and local loss $\lambda+\kappa$. 
    (b) Normalized relaxation time $\tau_{L, {\rm b}} \Delta_{\rm b}$ (c.f.~Eq.~(\ref{eq:boson_scaling})) for bosons, for various $\Gamma$  in the highly non-reciprocal limit $w = 1, \kappa = 0.999$ leading to a fixed localization length $\xi_{\rm loc} = 0.26 $. The EVec prediction $\tau_{\rm b}\Delta_{\rm b} \sim L/\xi_{\rm loc}$ gets the linear in $L$ scaling correct. However, as indicated by the dashed line in the figure, the EVec prediction diverges as $\xi_{\rm loc} \rightarrow 0$ (corresponding to $\kappa \rightarrow w$). Hence, any prefactor fitted to the actual scaling must be dramatically small, and will fail given even infinitesimal changes in $\kappa$.
    (c) Same as (b) but for fermions. Here, the linear-in-$L$ scaling of the EVec prediction is qualitatively wrong, with $\tau_{L, {\rm b}} \Delta_{\rm b}$ saturating with $L$.  
    As discussed in the main text, these strong deviations from the EVec prediction arise from an interference effect. 
    (d) $\tau_{L,{\rm f/b}}\Delta_{\rm f/b}$  as a function of local loss for $w = 1$, $\kappa = 0.999$, $\Gamma = 0.05$. Changing $\lambda$ does not change $\xi_{\rm loc}$ (the NHSE is unaffected), but strongly modifies 
    $\tau_{L,{\rm f/b}}$; this again strongly disagrees with the EVec prediction.  For large $\lambda$, we recover the behaviour of a reciprocal chain with the dissipative gap correctly predicting the relaxation time.
    } 
    \label{fig:perf_nr_plots}
\end{figure*}

\section{Quantum Hatano-Nelson model} 

The Hatano-Nelson model is a paradigmatic model exhibiting the NHSE. It can be realized in an unconditional, fully quantum setting using engineered dissipation  \cite{Metelmann_2015_PRX_nr_reseng}. The most natural realization is through the use of nearest-neighbour correlated loss \cite{AM_QHN_paper, liang2022_nhexpt}. The equation of motion for the density matrix using this setup reads (setting $\hbar = 1$)
\begin{align}\nonumber
     \partial_t \hrho
    & =
    -i\frac{w}{2}
    \sum_{j=1}^{L - 1}
    [
    \hc_{j+1}^\dagger \hc_j + \text{h.c.}, \hrho
    ]
    +\kappa
    \sum_{j=1}^{L-1} 
    \mathcal{D}[\hc_j-i\hc_{j+1}]\hrho
    \notag \\ \label{eq:L_HN}
    &\qquad  
    + 2   \sum_{j=1}^L \lambda_j \mathcal{D}[\hc_j]\hrho 
    +
    2\Gamma \sum_{j=1}^L\mathcal{D}[\hc_j^\dagger]\hrho
        \equiv
    \mathcal{L} \hrho,
\end{align}
where we consider an $L$-site lattice, open boundary conditions, 
and
$\lambda_j = \lambda + \kappa \left( \delta_{j,1} + \delta_{j,L} \right)$
\footnote{Note that the extra loss $\kappa$ maintains translational invariance on the chain: the local loss on each site is now exactly $\kappa + \lambda$.}.
This dissipative quantum realization of the Hatano-Nelson model has a variety of interesting properties, and is the subject of several recent studies \cite{wanjura_brunelli_nunnenkamp_2020, Wanjura_2020_PRL_correspondence_NH_ampl, Brunelli_2022_arxiv_restoration_nh_tops_ampl, Song_2019_PRL_chiral_damping, AM_QHN_paper, Anja_Hakan_PRA_2018}. Here $w$ is the coherent hopping strength, $\mathcal{D}[\hat{X}]\hrho \equiv \hat{X}\hrho \hat{X}^\dagger - \{\hat{X}^\dagger \hat{X}, \hrho\}/2$ is the usual dissipator, $\mathcal{L}$ is the Liouvillian superoperator, and $\hc_j$ are fermionic or bosonic annihilation operators for site $j$; we analyze both cases simultaneously. The degree of non-reciprocity can be tuned by varying the correlated loss $\kappa$. This becomes explicit in the equations of motion for the covariance matrix:
\begin{equation}\label{eqn:EOM}
\begin{aligned}
    &i \partial_t \langle \hat{c}_n^{\dag} \hat{c}_m \rangle 
    = 
    \\  
    & \sum_{j=1}^L
    \left( 
    (\mathbb{H}_{\rm HN} )_{mj} \langle \hat{c}_n^{\dag} \hat{c}_j \rangle 
    - 
    ( \mathbb{H}_{\rm HN}^{\dag} )_{jn} \langle \hat{c}_j^{\dag} \hat{c}_m \rangle  
    \right) 
    + 
    i 2\Gamma \delta_{nm}
    ,
\end{aligned}
\end{equation}
with $\mathbb{H}_{\rm HN}$ is the first-quantized Hatano-Nelson Hamiltonian
\begin{equation}
\begin{aligned}\label{eq:HN_Matrix}
\mathbb{H}_{\rm HN} \equiv 
&\sum_{j=1}^{L-1}
\left( 
\frac{w + \kappa}{2} | j + 1 \rangle \langle j |  
+  \frac{w - \kappa}{2} | j \rangle \langle j + 1 | 
\right) 
\\&-  i ( \kappa + \lambda \pm  \Gamma ) \sum_{j=1}^L|j\rangle \langle j|.
\end{aligned}
\end{equation}
Throughout, we will use the upper and lower sign $\pm$ for fermions and bosons respectively in addition to fixing $w \geq 0$ and $w> \kappa$, ensuring that the left-to-right hopping amplitude is stronger than right-to-left hopping. Note that the local loss proportional to $\kappa$ cannot be avoided if we want non-reciprocity. Without it, we could violate Pauli exclusion or the Heisenberg uncertainty principle \cite{Clerk_2010_RMP_noise_review}. The extra local loss $\lambda$ thus allows us to vary the degree of non-reciprocity without changing the total local loss. Incoherent pumping $\Gamma$ ensures the existence of a steady state with a finite density of particles with a non-trivial spatial profile \cite{AM_QHN_paper}. Finally, note that $\mathbb{H}_{\rm HN}$ is distinct from the no-jump Hamiltonian \cite{gardiner_zoller_2010}.

In what follows, we will be interested in the relaxation time of local observables. Given that the non-reciprocal hopping in Eq.~(\ref{eq:HN_Matrix}) favors rightward propagation and we wish to consider the effects of system size on relaxation dynamics, it is  natural to consider the occupation of the rightmost site $\langle \hat{n}_L(t) \rangle$. We define the normalized non-equilibrium population \begin{equation}\label{eq:norm_neq_popn}
    \delta n_L(t) = \frac{\left| \langle \hat{n}_L(t) \rangle - \langle \hat{n}_L(\infty) \rangle \right|}{ \langle \hat{n}_L(\infty) \rangle}
\end{equation}
and for concreteness assume that the chain starts in vacuum: $\delta n_L(0) = 1$. To define a local relaxation time $\tau_L$, we fix a relaxation threshold $e^{-1}$, and say that site $L$ has relaxed at time $\tau_L$ when $\delta n_L(\tau_L) = e^{-1}$. The essential physics we discuss is not specific to the local observable of interest, initial state or choice of reasonable relaxation threshold. See App.~\ref{app:other_initial_conditions} for a discussion of other initial conditions, and App.~\ref{app:greens_functions} for a discussion of the full time-dependent decay curve.  

\section{Exponentially large coefficients} 


To solve Eq.~(\ref{eqn:EOM}) and extract $\tau_L$, we first diagonalize $\mathbb{H}_{\rm HN}$. The eigenvalues $E_{\alpha}$ and biorthonormal left and right eigenvectors $\langle \psi^l(\alpha) | \psi^r(\beta) \rangle = \delta_{\alpha \beta}$ of the Hatano-Nelson model are well known \cite{OG_QHN_1, OG_QHN_2}. We have 
\begin{align}\label{eq:Eigenenergies}
    E_\alpha = J \cos k_\alpha-i(\kappa + \lambda \pm \Gamma),
\end{align}
and
\begin{align}\label{eq:Eigenvectors}
    \langle j|\psi^{r(l)}(\alpha) \rangle\equiv
    \psi_j^{r(l)} (\alpha) = 
    \sqrt{\frac{2}{L+1}}
    e^{(-) j/\xi_{\rm loc}} \sin k_\alpha j,
\end{align}
with
\begin{align}\label{eq:J_xi_def}
    J \equiv \sqrt{(w+\kappa)(w-\kappa)}
    ,
    \: 
    e^{1/\xi_{\rm loc}}
    \equiv
    \sqrt{\frac{w+\kappa}{w-\kappa}}.
\end{align}
Here, $k_\alpha = \alpha\pi/(L+1), \alpha = 1, \dots, L$ is a quantized standing-wave momentum. The parameter $J$ should be thought of as a renormalized hopping amplitude, whereas $\xi_{\rm loc}$ is not only the localization length of the eigenvectors but serves as a measure of non-reciprocity. We stress that fermions and bosons have the same eigenvectors, and the only difference between the two is the uniform decay rate
\begin{equation}
    \Delta_{\rm f/b} \equiv - {\rm Im} E_{\alpha} = \kappa + \lambda \pm \Gamma.
\end{equation}
Having diagonalized $\mathbb{H}_{\rm HN}$, we can use third quantization to also diagonalize the full Liouvillian \cite{Prosen_3Q, Prosen_2010, AM_2022_unpublished}, analogous to how one diagonalizes a second-quantized quadratic Hamiltonian from its single-particle data. The eigenvalues of $\mathcal{L}$ are fully determined by $E_{\alpha}$ and its eigenmodes are simply related to the left and right eigenvectors of $\mathbb{H}_{\rm HN}$ and the steady-state correlation matrix $(\mathbb{S})_{nm} \equiv \langle \hc^\dagger_n \hc_m \rangle_{\rm ss}$.


We can now use the spectral decomposition of $\mathbb{H}_{\rm HN}$ to formally express $\langle \hat{n}_m(t) \rangle$ as a sum over eigenvectors 
\begin{equation}\label{eqn:formal_soln}
\begin{aligned}
    &\langle \hat{n}_m(t) \rangle 
    =
    2 \Gamma  \sum_{j=1}^m \int_{0}^{t} dt' 
    \left|
    \langle m | e^{-i \mathbb{H}_{\rm HN}(t-t')} |j \rangle
    \right|^2
    \\
    = 
    &2 \Gamma  
    \sum_{j=1}^m
    \int_0^t
    dt'
    \left|
    \sum_{\alpha}
    \psi^r_m(\alpha) (\psi^l_j(\alpha))^* e^{-i E_\alpha (t-t')}
    \right|^2.
    \\
    &
\end{aligned}
\end{equation}
This solves Eq.~(\ref{eqn:EOM}) for the vacuum initial condition $\langle \hat{n}_m(0) \rangle = 0$ for all $m$, hence picking out the homogeneous part of the solution \footnote{See Eq.~(S27) for other initial conditions, where the inhomogeneous term nonzero. }. Eq.~(\ref{eqn:formal_soln}) provides a simple physical picture of the dynamics. At some time $ t' \leq t$, a pump bath attached to site $j$ injects a particle, which then propagates to a site $m$ in a time $t-t'$ contributing 
\begin{equation}
    P(m, j; t-t') \equiv |\langle m | e^{-i \mathbb{H}_{\rm HN}(t-t')} |j \rangle|^2
\end{equation}
to the density at that site. With each bath adding particles at a rate $2\Gamma$, and considering we started in vacuum at time $t = 0$, summing over all sites $j$ and integrating over all intermediate times $t'$ gives the total particle number on site $m$. Henceforth we will focus mainly on the rightmost site $m = L$.

With the propagator $\langle L | e^{-i \mathbb{H}_{\rm HN} t} | j \rangle$ written in its spectral representation, from Eq.~(\ref{eq:Eigenvectors}) it seems apparent that for strong non-reciprocity $\kappa \to w$, $\xi_{\rm loc} \ll 1$, only particles injected on the first site $j= 1$ which propagate to site $L$ matter, since in this limit the $e^{2(L-1)/\xi_{\rm loc}}$ factor in this term (coming from the exponential localization of the right eigenvectors) will dominate over all others.  To estimate $\tau_L$, one might then argue that we must compare this large term with $e^{- \Delta_{\rm f/b} t}$, the temporal decay stemming from the imaginary part of the mode energies. $\tau_L$ is then roughly the time it takes for the temporal decay factor to cancel the exponentially-large expansion coefficient. We would thus approximately have
\begin{align}\label{eqn:naive_relaxation}
\tau_{L, \rm f/b}
\stackrel{?}{\sim}
 \frac{L}{\xi_{\rm loc} \Delta_{\rm f/b}}.
\end{align}
This is essentially the argument put forward in Refs.~\cite{Haga_PRL_2021_LSE, Mori_PRL_2020_LSE} -- as we show explicitly in App.~\ref{app:3Q}, the exponentially-large terms in Eq.~(\ref{eqn:formal_soln}) are related to expansion coefficients in the basis of Liouvillian eigenmodes.

To check Eq.~(\ref{eqn:naive_relaxation}), in Fig.~\ref{fig:perf_nr_plots}b. and \ref{fig:perf_nr_plots}c. we plot $\tau_L$ in the large non-reciprocal limit $\kappa = 0.999 w$ giving $\xi_{\rm loc} = 0.26$ and, for $L = 100$, an expansion coefficient of $e^{2(L-1)/\xi_{\rm loc}} \sim 10^{330}$ \footnote{One might argue that a localization length $\xi_{\rm loc}$ much smaller than the lattice constant cannot reasonably be considered a physical quantity. In App.~\ref{app:greens_functions}, we show that Eq.~(\ref{eqn:naive_relaxation}) is wrong even for $\xi_{\rm loc} \sim O(1)$.}. In Fig.~\ref{fig:perf_nr_plots}d., we plot the dependence of the relaxation time on the additional local loss parameter $\lambda$. In each case, the EVec prediction of Eq.~(\ref{eqn:naive_relaxation}) falls short, albeit in different ways.

\begin{figure}[t]
    \centering
    \includegraphics[width=\linewidth]{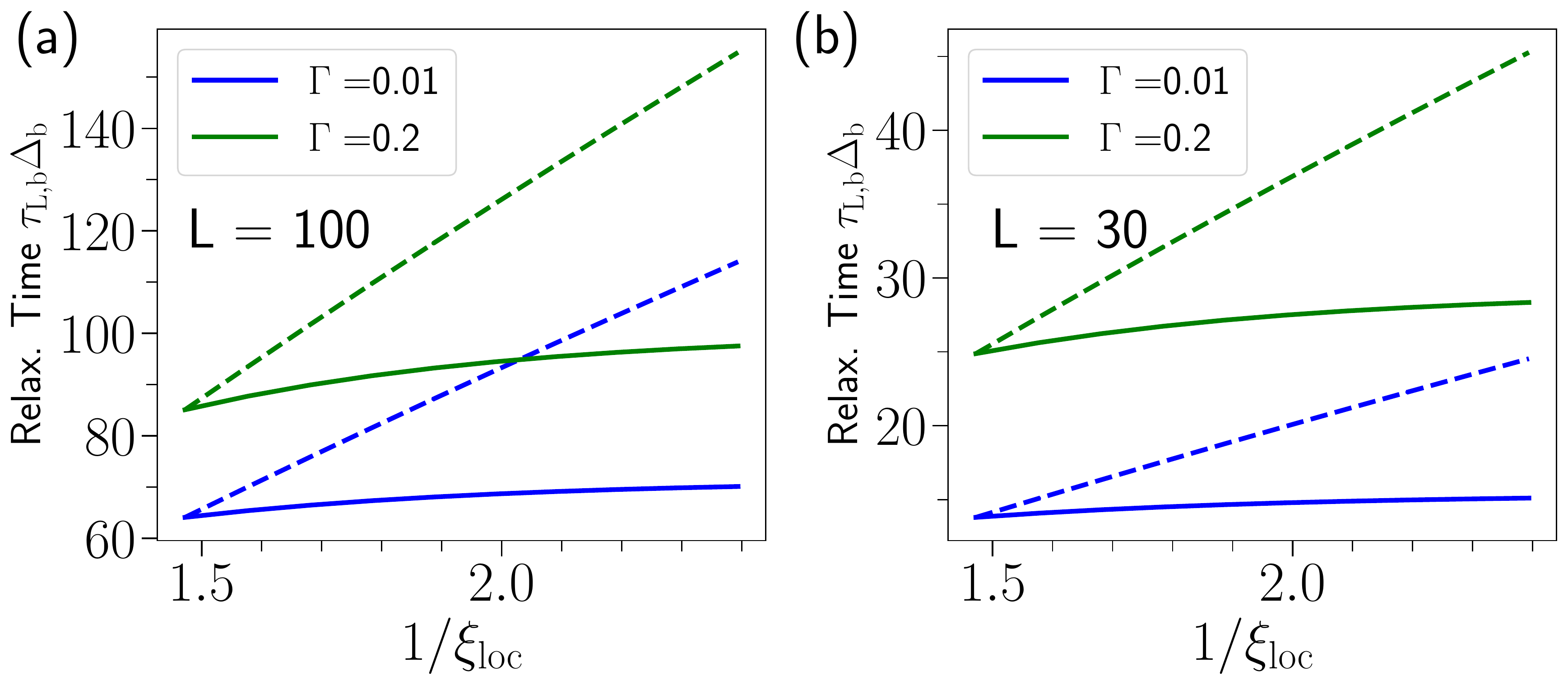}
    \caption{
    (a) Normalized relaxation time, as defined in Eq.~(\ref{eq:norm_neq_popn}), for the right-most site $\tau_{L, {\rm b}} \Delta_{\rm b}$ for bosons, with $w = 1, \Gamma = 0.01, 0.2, \lambda = 0, L = 100$ fixed, plotted against $1/\xi_{\rm loc}$. $\xi_{\rm loc}$ is varied by varying $\kappa$. Solid line indicates numerical result, and dashed line is obtained from the EVec. prediction. The prefactor for the EVec. prediction is fixed by setting the EVec. prediction to be equal to the numerical result for $\kappa = 0.9$.  We see that the relaxation times do not scale as $1/\xi_{\rm loc}$.
    (b) Similar to (a) but for $L=30$. In (a), $L \geq \xi_{\rm prop}$ for both cases, but here $L < \xi_{\rm prop}$ for $\Gamma = 0.01$. In both cases, the discrepancy remains.}
    \label{fig:boson_discrepancy_plot}
\end{figure}

First, for bosons (Fig.~\ref{fig:perf_nr_plots}b), Eq.~(\ref{eqn:naive_relaxation}) in fact gets the linear-in-$L$ scaling correct. The relaxation times for bosons show some dependence on $\Gamma$, which is independent of $\xi_{\rm loc}$, but one might argue that this is simply a prefactor difference. However, a more serious discrepancy appears 
when we consider the dependence of relaxation times of $\xi_{\rm loc}$. As an example, consider Fig.~\ref{fig:boson_discrepancy_plot}, where we fix all parameters except $\kappa$. Then, for a fixed chain length, varying $\kappa$ allows us to investigate the dependence of relaxation time on $\xi_{\rm loc}$, with $\xi_{\rm loc} \rightarrow 0$ when $\kappa \rightarrow w$. In this case, we observe that the EVec prediction diverges with $1/\xi_{\rm loc}$, while the true relaxation time saturates. 

Second, for fermions (Fig.~\ref{fig:perf_nr_plots}c),  Eq.~(\ref{eqn:naive_relaxation}) gets the scaling qualitatively wrong. The $\tau_L$ relaxation time only scales with system size up to a point, and ultimately saturates for large enough system sizes. Despite the NHSE, for large system sizes, fermions do not display the predicted linear-in-$L$ scaling of Eq.~(\ref{eqn:naive_relaxation}). 

Finally,  we find that  for a fixed $\xi_{\rm loc}$, the true relaxation times can vary drastically (Fig.~\ref{fig:perf_nr_plots}d). This is done by adding an local loss parameter which is completely independent of $\xi_{\rm loc}$. In this case, when $\lambda$ is sufficiently large, we even recover the usual dissipative gap scaling. We emphasize that at all points in this plot the NHSE is still present to the same degree quantified by $\xi_{\rm loc}$. However, anomalous relaxation has completely disappeared.


On the one hand, the argument leading to Eq.~(\ref{eqn:naive_relaxation}) seems obvious -- surely a drastic sensitivity to boundary conditions must show up in relaxation times. On the other, we have just demonstrated a range of ways in which Eq.~(\ref{eqn:naive_relaxation}) fails in a minimal model exhibiting the NHSE. While appealing, the argument leading to Eq.~(\ref{eqn:naive_relaxation}) is clearly misleading; we diagnose and correct the crucial error in what follows.

\section{Interference between eigenmodes} 

The failure of Eq.~(\ref{eqn:naive_relaxation}) makes it evident that we cannot simply estimate the magnitude of the propagator using the localization properties of the eigenvectors. To understand why, first note that Eq.~(\ref{eq:Eigenvectors}) implies 
\begin{equation}
     \langle m | e^{-i \mathbb{H}_{\rm HN} t} |j \rangle = e^{(m-j)/\xi_{\rm loc}- \Delta_{\rm f/b} t} \langle m | e^{-i \mathbb{H}_{J} t} |  j \rangle,
\end{equation}
where $\mathbb{H}_{J}$ is a \textit{reciprocal} nearest-neighbour tight-binding Hamiltonian with hopping amplitude $J$ under open boundary conditions. Focusing on the impact of this  previously-ignored second factor, we now work in the strongly non-reciprocal limit $\kappa \rightarrow w$ and set $\lambda = 0$, just as in Fig.~(\ref{fig:perf_nr_plots}) . In this limit, leftwards propagation is strongly suppressed. A particle which hits the right-most edge of the chain essentially cannot be reflected back and affect any other site; the boundary becomes irrelevant. Estimating the propagator by its fully translationally-invariant counterpart gives
\begin{align}\label{eq:Approx_P}
&P(L, j; t)
\approx
e^{2(L-j)/\xi_{\rm loc}-2 \Delta_{\rm f/b} t}
\left|
\int_{-\pi}^{\pi}
\frac{d k}{2\pi}
e^{i k(L-j)}
e^{-i J \cos k t}
\right|^2
\end{align}
and we show in App.~\ref{app:no_bounce} that this is in fact an excellent approximation in the regime of interest. 

Using Eq.~(\ref{eq:J_xi_def}) we deduce that although the first factor of Eq.~(\ref{eq:Approx_P}) is enormous for $j \ll L$ and strong reciprocity (i.e.~$\xi_{\rm loc} \ll 1$), the second factor in this limit can be very small,  as $J \to 0$. In fact, expanding to lowest order in $J$ and expressing all quantities in terms of $w$ and $\kappa$:
\begin{align}\nonumber
P(L,j;t)
&\approx 
\left(
J
e^{1/\xi_{\rm loc}}
\right)^{2(L-j)}
\frac{
e^{-2\Delta_{\rm f/b}t}
t^{2(L-j)}
}
{
4^{L-j}([L-j]!)^2
}
\\ \label{eq:Approx_P_Simplified}
&
=
\left(
w+\kappa
\right)^{2(L-j)}
\frac{
e^{-2\Delta_{\rm f/b}t}
t^{2(L-j)}
}
{
4^{L-j}([L-j]!)^2.
}
\end{align}
We see explicitly that the exponentially-large contribution from $\xi_{\rm loc}$ has been offset by the small renormalized hopping amplitude. 

The above dramatic cancellation is due to the interference between the eigenvectors of the Hatano-Nelson model, as in Eq.~(\ref{eq:Approx_P_Simplified}) we have summed over all $\alpha$ in the spectral decomposition of the propagator
\begin{equation}
    \langle L|e^{-i \mathbb{H}_{\rm HN}t}|j\rangle = \sum_{\alpha} \psi^r_L(\alpha) (\psi^l_j(\alpha))^* e^{-i E_\alpha t}.
\end{equation} 
Indeed, before Eq.~(\ref{eqn:naive_relaxation}) we incorrectly estimated the magnitude of $P(L,j;t)$ because we focused on a {\it single} eigenvector.  This incorrectly neglects the fact 
that for $J \to 0$, we have many modes with closely spaced eigenvalues $E_{\alpha}$, and the possibility of destructive interference when we sum their contributions. In App.~\ref{app:visualize}, we show this pictorially by plotting side-by-side the size of each term in the sum, along with their relative phases. While each term is exponentially large, the phase variation between terms leads to a near perfect cancellation in the sum over all eigenvectors.

While this explanation might seem mechanistic, it has a more general physical underpinning.  
This cancellation had to occur to enforce locality: a particle injected at $j \ll L$ cannot instantaneously propagate to $L$ with an exponentially-large amplitude, and hence cannot instantaneously know about the exponential modification of wavefunctions associated with the boundary condition.  We expand on this point in the next section, arguing that this reasoning also holds for intermediate times and is not limited to the Hatano-Nelson model. 

Despite demonstrating that the localization length plays no role in determining the relaxation time, it still does not explain why $\tau_L$ displays qualitatively different behavior for fermions and bosons nor why it is sensitive to the small incoherent pumping rate $\Gamma$. To uncover the origin of this effect, for simplicity consider the perfectly non-reciprocal case $\kappa = w$ where the dissipative gap takes the form $\Delta_{\rm f/b} = w \pm \Gamma$ and Eq.~(\ref{eq:Approx_P_Simplified}) is exact \footnote{Although this corresponds to an exceptional point pf $\mathbb{H}_{\rm HN}$, this structure is irrelevant here since the propagator $e^{-i \mathbb{H}_{\rm HN} t}$ is a smooth function of $\mathbb{H}_{\rm HN}$ and its parameters. Our conclusion holds for strong but not perfect non-reciprocity, just as in Fig.~(\ref{fig:perf_nr_plots})}. Fixing the distance $L-j$, a simple calculation reveals that $P(L,j;t)$ is a sharply-peaked function in the time domain, being maximal at
\begin{align}\label{eq:crit_t}
    t^{\max}_{L-j}
    \equiv
    \frac{L-j}
    {w\pm \Gamma} = \frac{L-j}
    {\Delta_{\rm f/b}}.
\end{align}
As shown explicitly in App.~\ref{app:greens_functions}, qualitatively it is correct to neglect the temporal integral in Eq.~(\ref{eqn:formal_soln}) and think about a particle injected at site $j$ as arriving at site $L$ at time $t^{\rm max}_{L-j}$ and contributing $P(L, j; t^{\rm max}_{L-j})$ to $\langle \hat{n}_L(t) \rangle$. 
We can hence interpret $t_{L-j}^{\rm max}$ as the effective time it takes for a particle to propagate a distance $L-j$.

\begin{figure}[t]
    \centering
    \includegraphics[width=\linewidth]{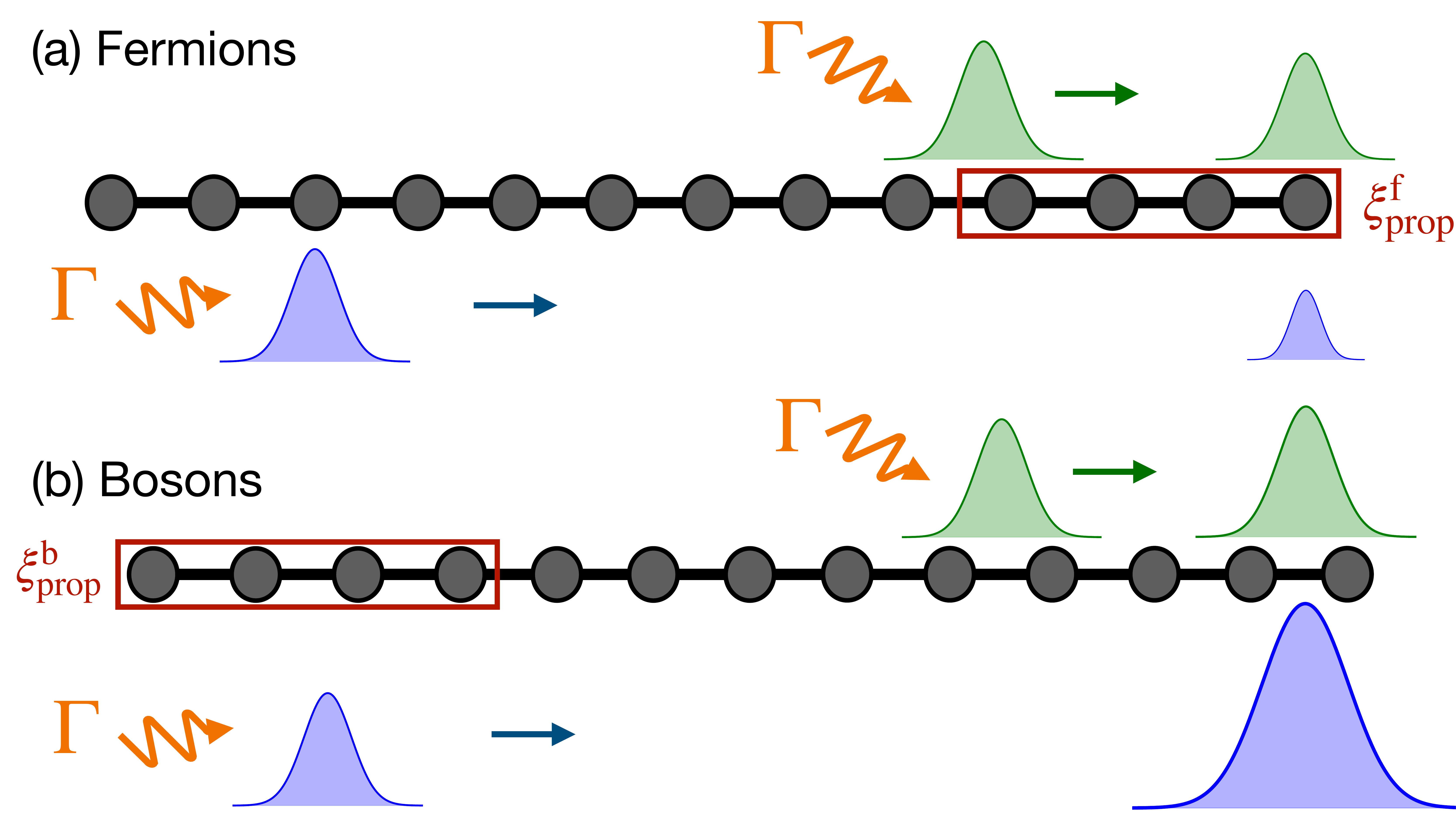}
    \caption{Schematic representation of particle propagation in the lattice. 
    (a) Fermions injected into the chain are de-amplified as they propagate to the right. Consequently, the primary contributions to the final site $L$ come from a small range of nearby sites within a distance $\xi^{\rm{f}}_{\rm prop}$ of the end of the chain. 
    (b) Bosons injected into the chain are amplified as they propagate to the right.
    Consequently, the primary contributions to the final site $L$ come from a small range of sites within a distance $\xi^{\rm{b}}_{\rm prop}$ of the start of the chain
    These processes set the decay time $\tau_{L}$ and the size of their contributions are by $w, \Gamma$, and as such are independent of the localization length}; see Eqs.~(\ref{eq:fermion_scaling})-(\ref{eq:boson_scaling}), and Eq.~(\ref{eq:P_max}), and accompanying discussion. 
    \label{fig:prop_schematic}
\end{figure}

Surprisingly, depending on the type of particle, the amplitude $P(L,j;t) = |\langle L | e^{-i\mathbb{H}_{\rm eff} t}|j \rangle|^2$ evaluated at its peak has a strikingly different spatial behavior
\begin{align}\label{eq:P_max}
    P(L,j; t^{\max}_{L-j})
    \propto
    \left(
    \frac{w}{w\pm \Gamma}
    \right)^{2(L-j)}
    \equiv
    e^{\mp(L-j)/\xi_{\rm prop}^{\rm f/b}}
    .
\end{align}
As a function of distance $L-j$, we thus have exponential \textit{growth} (amplification) for bosons, whereas for fermions we have the opposite: exponential \textit{decay} (attenuation). The length scale controlling this behavior is not $\xi_{\rm loc}$ but rather $(\xi_{\rm prop}^{\rm f/b})^{-1} = \mp 2\ln(w/(w\pm \Gamma)) >0$. For fermions, this attenuation means that only particles injected within a length $\xi_{\rm prop}^{\rm f}$ of the left site $L$ can reach it; adding additional sites on the leftmost side of the chain effectively does not change $\langle \hat{n}_L \rangle_{\rm ss}$. Using Eq.~(\ref{eq:crit_t}) and assuming $L\gg \xi_{\rm prop}^{\rm f}$, we must then only wait a time 
\begin{align}\label{eq:fermion_scaling}
    \tau_{L, \rm f}
    \sim
   t^{\max}_{\xi_{\rm prop}^{\rm f}}
    \approx
    \frac{\xi_{\rm prop}^{\rm f}}{\Delta_{\rm f}}
\end{align}
before all particles in the steady state are accounted for, explaining the saturation we see in Fig.~(\ref{fig:perf_nr_plots}c). For bosons however the amplification physics implies that, for {\color{red} $L \gg \xi_{\rm prop}^{\rm b}$}, particles on the first site contribute the \textit{most} to the occupation on the last site. Consequently, we must wait for a particle injected on the first site to propagate the length of the chain leading to 
\begin{align}\label{eq:boson_scaling}
    \tau_{L, \rm b}
    \sim
    t^{\max}_{L-1}
    \approx
    \frac{L}{\Delta_{\rm b}}.
\end{align}
This analysis is corroborated by Fig.~(\ref{fig:perf_nr_plots}b) of the main text, and Fig.~(S4) in App.~\ref{app:greens_functions}.  At first glance, this seems contradictory -- the only difference between the two types of particles enters as a sign in Eq.~(\ref{eq:P_max}), which disappears as $\Gamma \rightarrow 0$. Yet, Eqs.~(\ref{eq:boson_scaling}) and (\ref{eq:fermion_scaling}) paint a very different picture of relaxation for bosons and fermions. This is another main result of this work: amplifying and decaying dynamics have markedly different relaxation behavior when $L \gg \xi_{\rm prop}^{f/b}$. The intuitive picture of propagation for both types of particles are schematically depicted in Fig.~(\ref{fig:prop_schematic}). However, the amplifying and decaying dynamics are heavily dependent on length scale, and when $L \ll \xi_{\rm prop}^{f/b}$, the relaxation behavior of both types of particles are the same. The exact relaxation curves in this small $L$ limit, while tangential to our main result, are also carefully worked out in App.~\ref{app:greens_functions}.

\section{Generic tight-binding Hamiltonians}

It is natural to ask whether the relaxation behaviour of the quantum Hatano-Nelson model is generic. Suppose now that we have realized a short-ranged non-Hermitian tight-binding Hamiltonian $\mathbb{H}_{\rm eff}$ exhibiting the NHSE using engineered Markovian dissipation, in the sense that the dynamics of the covariance matrix obey Eq.~(\ref{eqn:EOM}), perhaps with a different inhomogeneous term. Under open boundary conditions, the eigenvectors and eigenvalues of $\mathbb{H}_{\rm eff}$ are $\psi^{r(l)}_{\boldsymbol{m} \boldsymbol{\sigma}}(\alpha) \sim e^{(-) \boldsymbol{m}/\xi_{\alpha}}$ and $E_{\alpha}$ respectively, where $\boldsymbol{m}$ denotes a site of an arbitrary-dimensional lattice, $\boldsymbol{\sigma}$ indexes all other degrees of freedom and $\xi_{\alpha}$ is the localization length of mode $\alpha$. The propagator in the position basis
\begin{equation}
    \langle \boldsymbol{m} \boldsymbol{\sigma}|
e^{-i \mathbb{H}_{\rm eff} t} 
| \boldsymbol{j} \boldsymbol{\sigma'} \rangle 
= 
\sum_{\alpha} 
\psi^{r}_{\boldsymbol{m} \boldsymbol{\sigma}}(\alpha)
(\psi^{l}_{\boldsymbol{j} \boldsymbol{\sigma'}}(\alpha))^*
e^{- iE_{\alpha} t}
\end{equation}
can be used to determine any local observable, just as in Eq.~(\ref{eqn:formal_soln}). How can we generically conclude that the eigenvectors will interfere and render the relation dynamics essentially insensitive to $\xi_{\alpha}$?

Ultimately, the answer is locality. To be more precise, consider expanding the propagator to first order in $t$, 
\begin{align}\nonumber
    &\langle \boldsymbol{m} \boldsymbol{\sigma}|
e^{-i \mathbb{H}_{\rm eff} t} 
| \boldsymbol{j} \boldsymbol{\sigma'} \rangle 
= 
\sum_{\alpha} 
\psi^{r}_{\boldsymbol{m} \boldsymbol{\sigma}}(\alpha)
(\psi^{l}_{\boldsymbol{j} \boldsymbol{\sigma'}}(\alpha))^*
e^{- iE_{\alpha} t}
\\ \nonumber
&=
\delta_{\boldsymbol{m},\boldsymbol{j}}
\delta_{\boldsymbol{\sigma}, \boldsymbol{\sigma'}}
-
i t
\sum_{\alpha} 
E_{\alpha}
\psi^{r}_{\boldsymbol{m} \boldsymbol{\sigma}}(\alpha)
(\psi^{l}_{\boldsymbol{j} \boldsymbol{\sigma'}}(\alpha))^*
+
\mathcal{O}(t^2)
\\
&=
\delta_{\boldsymbol{m},\boldsymbol{j}}
\delta_{\boldsymbol{\sigma}, \boldsymbol{\sigma'}}
-
i t 
(\mathbb{H}_{\rm eff})_{(\boldsymbol{m}, \boldsymbol{\sigma}), (\boldsymbol{j}, \boldsymbol{\sigma'})}
+
\mathcal{O}(t^2)
\end{align}
In going from the second line to the third, we have made use of the spectral decomposition of $\mathbb{H}_{\rm eff}$. This simple calculation already highlights the crucial interference effect: summing over the localized eigenmodes $\alpha$ gives the matrix element of a \textit{short-ranged local Hamiltonian}, and not something that is exponentially large. Continuing this line of reasoning, for sites $\boldsymbol{m}$ and $\boldsymbol{j}$ located far enough in the bulk the propagator will be completely insensitive to boundary conditions to some large order in $t$. Going beyond this perturbative-in-$t$ reasoning, one would argue that for a sufficiently local Hamiltonian, the existence of a Lieb-Robinson velocity \cite{Lieb_1972_LR_bound} implies that there exists a timescale over which the presence or absence of boundaries is essentially irrelevant. 

The spectral decomposition however is of course exact for all times; the eigenvectors and eigenvalues are after all effectively infinitely-temporally resolved objects.  The only way to obtain recover both locality and a macroscopic number of spatially-localized edge modes is a large cancellation between eigenvectors at intermediate times; the exponential sensitivity to boundary conditions is not an instantaneous effect. In a sense, although the NHSE drastically changes the spectrum and eigenvectors of $\mathbb{H}_{\rm eff}$, \textit{they must change together in just the right way to render the local propagation physics effectively the same}. 

\section{Conclusion}

We have demonstrated that while non-Hermitian models exhibiting the NHSE can exhibit anomalous relaxation times, this is not in general simply determined by the localization length of eigenvectors.  This can be understood as resulting from an interference effect, and more generally on considerations involving locality (which limits the sensitivity to changes in boundary conditions).  While our conclusions are based on detailed studies of the quantum Hatano-Nelson model (and two additional setups in App.~\ref{app:other_models}), we believe that they are generic to any local model. Our work also highlights and emphasizes how there is a timescale associated with the presence of boundaries in models exhibiting the NHSE. Future works exploring how the effects of interactions impact this analysis would be extremely interesting, as would attempting to place quantitative bounds on how boundaries affect finite-time dynamics.  


\section{Acknowledgements}

This work was supported by the Air Force Office of Scientific Research MURI program under Grant No. FA9550-19-1-0399, and by the Simons Foundation (Grant No. 669487, A. C.).

\appendix

\section{Visualizing interference}\label{app:visualize}

\begin{figure}[htpb]
    \centering
    \includegraphics[width=0.99\linewidth]{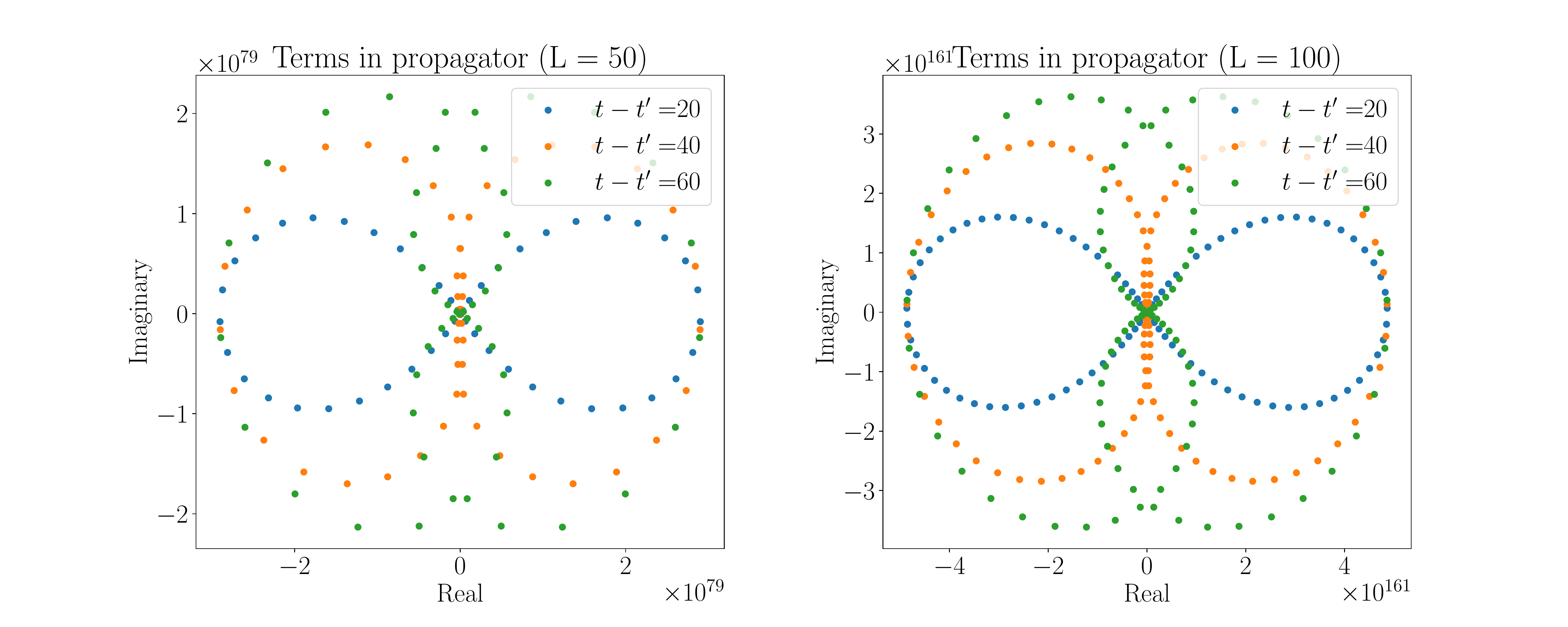}
    \caption{Plot of the terms $\psi^r_L(\alpha) (\psi^l_1(\alpha))^* e^{-i E_\alpha (t-t')}$ in the complex plane for  $w = 1, \kappa = 0.999$ and (a) $L = 50$, (b) $L = 100$. Note that in both cases the axes are scaled by a large exponential factor, due to the exponential largeness of individual terms. We emphasize that while individual terms are exponentially large, their sum may not be, due to the presence of a macroscopic number of exponentially large terms with differing phases.}
    \label{fig:visualize_interference}
\end{figure}

In the main text, we have argued that one cannot pick out any single exponentially large term in Eq.~(\ref{eqn:formal_soln}) as an estimate of its size. The reason for that is two-fold: (1) There is an extensive number of terms in the sum that are all exponentially large. (2) The phases of the terms in the sum are distributed across the complex plane. Here we provide an explicit plot of the terms in the sum, namely, 
\begin{equation}
\psi^r_L(\alpha) (\psi^l_1(\alpha))^* e^{-i E_\alpha (t-t')}    
\end{equation}
for different values of $t - t'$ and parameters in the highly non-reciprocal regime, $w = 1, \kappa = 0.999, L = 50, 100$, in order to visually demonstrate that this statement. As observed in Fig.~\ref{fig:visualize_interference}, there are a large number of terms that are all exponentially large, and these terms are distributed across different phases, and hence have the possibility of cancelling out. Note that the density of exponentially large terms increases with $L$, which is a signature of the macroscopic localization associated with the NHSE.

\section{No bounce approximation}\label{app:no_bounce}

The goal of this section is to show that Eq.~(\ref{eq:Approx_P}) of the main text, which we refer to as the \textit{no-bounce approximation} for reasons which will become apparent, is in fact an excellent approximation for our model in the highly non-reciprocal regime. The equations of motion for the retarded Green's function $G(m, j; t) = \langle m | e^{- i \mathbb{H}_{\rm HN} t} | j \rangle$ are given by \cite{bruus_flensberg_2020}
\begin{widetext}
\begin{equation}\label{app_eq:G_EOM}
    \begin{aligned}
    i \partial_t G(m, j; t) 
    = \frac{1}{2} J e^{1 / \xi_{\rm loc}} G(m - 1, j; t) + \frac{1}{2} J e^{- 1 / \xi_{\rm loc}} G(m + 1, j; t) 
    - i \Delta_{\rm f/b} G(m, j; t), 
    \end{aligned}
\end{equation}
\end{widetext}
where $J, \xi_{\rm loc}, \Delta_{\rm f/b}$ are the renormalized hopping, localization length, and dissipative gap respectively; each is defined in the main text (see Eq.~(\ref{eq:J_xi_def}) and the paragraph directly below it). The initial condition is given by $G(m, j, 0) = \delta_{mj}$.

As a first step, we consider $G_{\infty}(m, j; t)$, which is the solution to the above EOM for an infinite lattice. For an infinitely long chain, there are no boundaries and thus no quantization condition on the momentum. Solving the equations of motion yields,
\begin{widetext}
\begin{equation}\label{app_eq:G_inf_SM}
\begin{aligned}
    G_{\infty}(m, j, t) 
    &= e^{(m - j)/ \xi_{\rm loc}} e^{- \Delta_{\rm f/b} t} \int_{- \pi}^{\pi} \frac{dk}{2 \pi} e^{ik(m - j)} e^{-i J \cos k t} 
    \\&\equiv e^{(m - j)/ \xi_{\rm loc}} e^{- \Delta_{\rm f/b} t} G_{\infty}^J(m - j; t)
\end{aligned}    
\end{equation}
\end{widetext}
where 
\begin{equation*}
    G_{\infty}^J(m - j; t) \equiv \int_{- \pi}^{\pi} \frac{dk}{2 \pi} e^{ik(m - j)} e^{-i J \cos k t} = G_{\infty}^J(|m - j|; t)
\end{equation*}
is the Green's function of an infinite \textit{reciprocal} chain with hopping $J$. 

The EOM for a finite-sized system under open boundary conditions (OBC) is exactly the same as Eq.~(\ref{app_eq:G_EOM}); only the boundary terms $G_{\rm OBC}(j,L+1; t) = G_{\rm OBC}(j,0;t) = G_{\rm OBC}(L+1,m;t) = G_{\rm OBC}(0,m)$ are different. Thus, by linearity, we can construct the solution for a finite-sized chain with OBC from a linear combination of solutions for the infinite chain. The full solution reads \cite{AM_QHN_paper}
\begin{widetext}
\begin{equation}
\begin{aligned}
G_{OBC}(m, j; t) 
= e^{(m - j)/ \xi_{\rm loc}} e^{- \Delta_{\rm f/b} t} 
 \sum_{b = -\infty}^{\infty} \left(   G_{\infty}^J(m - j + 2 (L + 1) b; t) \right.
 \left. - G_{\infty}^J(m + j + 2 (L + 1) b; t) \right).
\end{aligned}
\end{equation}
\end{widetext}
We stress that each term in the sum indexed by $b$ can be interpreted as the particle ``bouncing" off either the left or right edge. 

To recover the more standard expression written in terms of the eigenvectors of the $L$-site chain, we can use the integral representation of $G^J_\infty(m-j;t)$ and Poisson's summation formula (or equivalently a standard representation of the Dirac comb) and obtain
\begin{widetext}
\begin{align} \nonumber
    &G_{OBC}(m, j; t) 
= e^{(m - j)/ \xi_{\rm loc}} e^{-\Delta_{\rm f/b}t}
\frac{1}{2(L+1)}
\\ \nonumber
&\qquad \times
\Biggl(
\sum_{\alpha = -\infty}^{\infty}
\int_{-\pi}^\pi
\frac{d k}{2(L+1)}
e^{i k(m-j) -i J \cos k t}\delta\left(k- \frac{\alpha \pi}{L+1}\right)
-
\sum_{\alpha = -\infty}^{\infty}
\int_{-\pi}^\pi
\frac{d k}{2\pi}
e^{i k(m+j) -i J \cos k t}\delta\left(k- \frac{\alpha \pi}{L+1}\right)\Biggr).
\end{align}
\end{widetext}
Integrating over $k$ gives the quantization over 
\begin{equation*}
    k_{\alpha} = \frac{\pi \alpha}{L+1},
\end{equation*}
\bigskip
and thus
\begin{widetext}
\begin{align}
    G_{OBC}(m,j;t)
    &=
    \frac{2 e^{(m - j)/ \xi_{\rm loc}} e^{-\Delta_{\rm f/b}t}}{L+1}
    \times
    \sum_{k_\alpha = 1}^L
    e^{-i J \cos k_\alpha t}
    \sin k_\alpha m \sin k_{\alpha} j
\end{align}
\end{widetext}
is equivalent to the Green's function computed from the spectral data (Eq.~(\ref{eq:Eigenenergies})-(\ref{eq:Eigenvectors}) of the main text) and the eigenenergies $E_\alpha = J \cos k_\alpha - i \Delta_{\rm f/b}$

Finally, we consider the fully non-reciprocal limit explored in most of the main text. In this limit, $J \rightarrow 0$. Since $G_{\infty}^J(d, t) \sim J^{|d|}$, to the lowest order in $J$, we only need to keep the $b = 0$ term in the sum. This justifies and yields the \textit{no-bounce approximation}, 
\begin{equation}\label{app_eq:G_OBC_J_SM}
\begin{aligned}
G_{OBC}(m, j; t) 
\simeq &e^{(m - j)/ \xi_{\rm loc}} e^{- \Delta_{\rm f/b} t} 
 G_{\infty}^J(m - j; t),
\end{aligned}
\end{equation}
which is applied in Eq.~(\ref{eq:Approx_P}) of the main text.

\section{NHSE implies the Liouvillian Skin Effect}\label{app:3Q}

The goal of this section is to show that the eigensystem of the effective Hamiltonian $\mathbb{H}_{\rm eff}$ fully characterize the eigenmodes of the Liouvillian $\mathcal{L}$. To do so, we first provide a brief introduction to the formalism of \textit{third quantization}, first introduced by Prosen \cite{Prosen_3Q} and subsequently used by Prosen and Seligman \cite{Prosen_2010}. The details of third quantization recalled in this section are further expanded upon in \cite{AM_2022_unpublished}.

First, we can identify operators as elements of a Hilbert space by writing $\hat{X} \rightarrow | \hat{X} \rangle \rangle$, and introducing the inner product between two operators 
\begin{equation}
    \langle \langle \hat{Y} | \hat{X} \rangle \rangle \equiv \mbox{Tr} (\hat{Y}^{\dag} \hat{X}).
\end{equation}

The left and right eigenmodes of the Liouvillian superoperator $\mathcal{L}$ can now be written 
\begin{equation}\label{app_eq:LR_def}
    \hat{\mathcal{L}} | r_{\alpha} \rangle \rangle = \lambda_{\alpha} | r_{\alpha} \rangle \rangle
    \qquad \hat{\mathcal{L}}^{\dag}  | l_{\alpha} \rangle \rangle =  \lambda_{\alpha}^*  | l_{\alpha} \rangle \rangle,
\end{equation}
where $\alpha$ indexes the eigenmodes, the eigenvalues $\lambda_{\alpha}$ are in general complex, and we have introduced the hat notation to $\hat{\mathcal{L}}$ to emphasize that it is an operator in this space. Our goal is to show that for the quadratic problem at hand, a subset of $\lambda_{\alpha}$ and $|l_\alpha\rangle$ which are relevant to the observables of interest can be expressed in terms of the eigenvalues and eigenvectors of the dynamical $\mathbb{H}_{\rm eff}$ and steady-state correlation $\mathbb{S}_{\rm ss}$ matrices. By effective Hamiltonian, we mean the dynamical matrix which controls the equations of motion of the two-point correlation matrix. In the main text, $\mathbb{H}_{\rm eff} = \mathbb{H}_{\rm HN}$.

The most general model we consider are $U(1)$-symmetric Lindbladians which have quadratic Hamiltonians and jump operator which are linear in $\hc_m$ and $\hc_n^\dagger$. That is, 
the action of $\hat{\mathcal{L}}$ on an arbitrary operator
 $\hat{X}$ is
 \begin{widetext}
\begin{equation}
    \begin{aligned}
    \hat{\mathcal{L}} \hat{X} =
    & -i \sum_{nm} \mathbb{H}_{nm} [ \hat{c}_n^{\dag} \hat{c}_m, \hat{X} ]
     + \sum_{nm} \mathbb{L}_{nm} \left( \hat{c}_m \hat{X} \hat{c}_n^{\dag} - \frac{1}{2} \{ \hat{c}_n^{\dag} \hat{c}_m, \hat{X} \}\right)
     + \sum_{nm} \mathbb{P}_{nm} \left( \hat{c}_n^{\dag} \hat{X} \hat{c}_m - \frac{1}{2} \{ \hat{c}_m \hat{c}_n^{\dag} , \hat{X} \}\right),
    \end{aligned}    
\end{equation}
\end{widetext}
where $\mathbb{H}$ describes the underlying closed system Hamiltonian dynamics and $\mathbb{L}, \mathbb{P}$ are matrices describing the action of the loss and gain dissipators. Note that to be in Lindblad form, $H$ must be Hermitian, whereas $L, P$ are Hermitian and positive semi-definite \cite{AM_QHN_paper}.

The adjoint $\hat{\mathcal{L}}^{\dag}$ is defined in the usual manner via the inner product be defined via the $\langle \langle \hat{\mathcal{L}}^{\dag} \hat{Y} | \hat{X} \rangle \rangle
    = \mbox{Tr} (\hat{Y}^{\dag} \hat{\mathcal{L}} \hat{X} )
    = \langle \langle  \hat{Y} |  \hat{\mathcal{L}} \hat{X} \rangle \rangle$. Using the cyclic property of the trace, we obtain 
    
\begin{widetext}
\begin{equation}
\begin{aligned}
    \hat{\mathcal{L}}^{\dag} \hat{Y} =
    & - i \sum_{nm} \mathbb{H}_{nm} [ \hat{c}_m^{\dag} \hat{c}_n, \hat{Y} ] 
     + \frac{1}{2} \sum_{nm} \mathbb{L}_{nm} \left( \hat{c}_n^{\dag} [ \hat{Y}, \hat{c}_m ] + [\hat{c}_n^{\dag}, \hat{Y} ] \hat{c}_m \right) 
     + \frac{1}{2} \sum_{nm} \mathbb{P}_{nm} \left( \hat{c}_m [ \hat{Y}, \hat{c}_n^{\dag} ] + [\hat{c}_m, \hat{Y} ] \hat{c}_n^{\dag} \right) 
\end{aligned}    
\end{equation}
\end{widetext}

Since we are interested in two-point correlation functions, we now let $\hat{Y} = \hat{c}_a^{\dag} \hat{c}_b$. Using the commutation relation $[\hat{c}_i, \hat{c}_j^{\dag}] = \delta_{ij}$ for bosons, or the anticommutation relation $\{\hat{c}_i, \hat{c}_j^{\dag}\} = \delta_{ij}$ for fermions, to get 
\begin{equation}\label{app_eq:LdagEOM}
\begin{aligned}
    \hat{\mathcal{L}}^{\dag} (\hat{c}_a^{\dag} \hat{c}_b) =
    & - i \sum_n (\mathbb{H}_{\rm eff})_{bn} \hat{c}_a^{\dag} \hat{c}_n
    + i \sum_n (\mathbb{H}_{\rm eff}^{\dag})_{na} \hat{c}_n^{\dag} \hat{c}_b + \mathbb{P}_{ab},
\end{aligned}    
\end{equation}
where 
\begin{equation}
\begin{aligned}
    \mathbb{H}_{\rm eff} = \mathbb{H} - \frac{i}{2} (\mathbb{L} \pm \mathbb{P}),
\end{aligned}    
\end{equation}
where as in the main text, the $+$ sign is for fermions and the $-$ sign is for bosons.

We will use Eq.~(\ref{app_eq:LdagEOM}) to obtain the left eigenmode of the Liouvillian in terms of the left eigenvectors of $\mathbb{H}_{\rm eff}$. Suppose we can diagonalize the non-Hermitian $\mathbb{H}_{\rm eff}$ as 
\begin{equation}
\begin{aligned}
    (\mathbb{H}_{\rm eff})_{ab} = \sum_{\alpha} E_{\alpha} \psi_a^r(\alpha) [\psi_b^l(\alpha)]^*,
\end{aligned}    
\end{equation}
where $ E_{\alpha}$ are the eigenenergies indexed by $\alpha$, and $\psi_a^{r/l}(\alpha) =  \langle a|\psi^{r/l}(\alpha) \rangle$. Note that $E_{\alpha}$ are also often referred to in the literature as the rapidities of the Lindbladian.

So far, we have not taken full advantage of the third quantization, having only used it to simplify the derivation of the operator equation of motion. To fully appreciate the idea, let us derive the eigenmodes of the full Liouvillian. To that end, consider the quantity 
\begin{equation}\label{app_eq:lefteig}
\begin{aligned}
    \hat{l}_{\alpha, \beta} \equiv \sum_{ab} \psi_{a}^l(\alpha) [\psi_b^l(\beta)]^* \hat{c}^{\dag}_a \hat{c}_b - S_{\alpha \beta},
\end{aligned}    
\end{equation}
where 
\begin{equation}
\begin{aligned}
    S_{\alpha \beta} \equiv -i \sum_{ab} \frac{\psi_{a}^l(\alpha) \mathbb{P}_{ab} [\psi_b^l(\beta)]^*}{E_{\alpha} - E_{\beta}^{*}}.
\end{aligned}    
\end{equation}

Inserting Eq.~(\ref{app_eq:lefteig}) into Eq.~(\ref{app_eq:LdagEOM}), and using the fact that the adjoint Liouvillian annihilates the identity $\hat{\mathcal{L}}^\dagger(S_{\alpha \beta}) = 0$ we find 
\begin{equation}
\begin{aligned}
    \hat{\mathcal{L}}^{\dag} (\hat{l}_{\alpha, \beta}) = - i (E_{\beta}^* - E_{\alpha}) \hat{l}_{\alpha, \beta}.
\end{aligned}    
\end{equation}
By comparing against the definition of the left eigenmode in Eq.~(\ref{app_eq:LR_def}), $\hat{l}_{\alpha, \beta}$ is a left eigenmode of $\hat{\mathcal{L}}$ with eigenvalue $i (E_{\beta} - E_{\alpha}^*)$. It can be shown that $S_{\alpha\beta}$ is in fact related to the normal-ordered steady-state correlation matrix $(\mathbb{S}_{\rm ss})_{ab} = \langle \hc^\dagger_a \hc_b\rangle_{\rm ss}$ via $S_{\alpha \beta} = \langle \psi^l(\alpha) | \mathbb{S} |\psi^r(\beta) \rangle $ \cite{AM_QHN_paper}.

We observe that the a set of left eigenvectors of the \emph{full Liouvillian} can be written in terms of the left eigenvectors $\psi^l(\alpha)$ of the \emph{effective Hamiltonian}. An analogous derivation proceeds for the right eigenmodes using $\hat{\mathcal{L}}(\hat{c}^{\dag}_a \hat{c}_b)$. In fact, as shown by Prosen \cite{Prosen_3Q} and Prosen and Seligman \cite{Prosen_2010}, this holds for all eigenvectors and eigenvalues; they are completely determined by $\mathbb{H}_{\rm eff}$ and $\mathbb{S}_{\rm ss}$ (see also Ref.~\cite{AM_2022_unpublished}). 
The conclusion is that if $H_{\rm eff}$ exhibits the NHSE, then the Lindbladian exhibits the LSE. The upshot is that while Eq.~(\ref{eqn:formal_soln}) is written over a sum of eigenvectors of $\mathbb{H}_{\rm eff}$, it also formally corresponds to an expansion of the density matrix in its basis of eigenvectors.

\section{The Green's functions of the Hatano-Nelson model}\label{app:greens_functions}

\begin{figure}[htpb]
    \centering
    \includegraphics[width=0.99\linewidth]{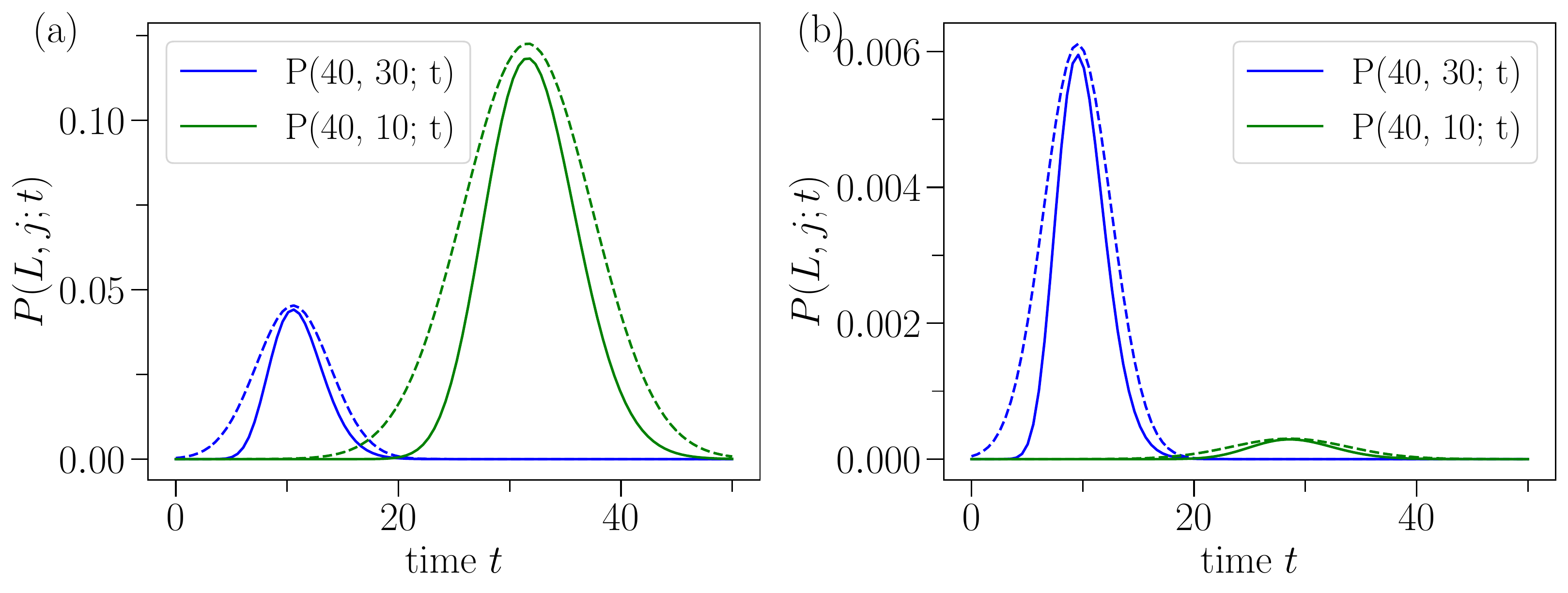}
    \caption{Plot of $P(40, 30; t)$ and $P(40, 10; t)$ (see Eq.~(\ref{eq:Approx_P_SM}))for bosons (a) and fermions (b) for parameters $w = 1, \kappa = 0.999, \Gamma = 0.05$. The peaks are located at $t_c = \frac{L - j}{\Delta_{f/b}}$, and the functions are roughly centered around these peaks. Notice that the area under the curve is much larger for $P(40, 10; t)$ than $P(40, 10; t)$ for bosons, whereas the opposite is true for fermions. This is due to the amplification dynamics of the non-reciprocal chain. Dashed lines depict the Gaussian approximation for $P$ (see Eq.~(\ref{eq:gaussian_sigma_SM}) and surrounding discussion). Note that the behavior in this figure is `typical' in the sense that the outlined features are generic when $L - j \gg \xi_{\rm prop}$.}
    \label{fig:P_fermion_boson}
\end{figure}

The goal of this section is to elucidate the behavior of the propagation probability $P(L, j; t)$ (Eq.~(\ref{eq:Approx_P})). For clarity, we reiterate the required definitions from the main text. The propagation probability from site $j$ to $L$ is,
\begin{align}\label{eq:Approx_P_SM}
&P(L, j; t)
\approx
e^{2(L-j)/\xi_{\rm loc}-2 \Delta_{\rm f/b} t}
\left|
\int_{-\pi}^{\pi}
\frac{d k}{2\pi}
e^{i k(L-j)}
e^{-i J \cos k t}
\right|^2,
\end{align}
where 
\begin{align}\label{eq:J_xi_def_SM}
    J \equiv \sqrt{(w+\kappa)(w-\kappa)}
    ,
    \: 
    e^{1/\xi_{\rm loc}}
    \equiv
    \sqrt{\frac{w+\kappa}{w-\kappa}},
\end{align}
with $\xi_{\rm loc}$ being the localization length, $J$ the renormalized hopping and $w$ the bare hopping of the underlying tight-binding chain. $\kappa$ describes correlated loss, $\Gamma$ describes pumping, and $\Delta_{\rm f/b} = \kappa \pm \Gamma$ is the dissipative gap.

Some typical examples of $P$ are plotted in  Fig.~\ref{fig:P_fermion_boson}. The three relevant quantities we will use to physically interpret these functions are speed, height, and width. Understanding the behavior of these Green's functions is essential to understanding how the relaxation times scale in the model. We will first extract these quantities, before summarizing their physical interpretations and consequences for the scaling behavior of relaxation times.

First, for strong non-reciprocity $\xi_{\rm loc} \ll 1$, we can simplify Eq.~(\ref{eq:Approx_P_SM}) by expanding to lowest order in $J$. In terms of $w, \kappa$, the simplified expression is given by Eq.~(\ref{eq:Approx_P_Simplified}) of the main text, 
\begin{align}\label{eq:Approx_P_Simplified_SM}
P(L,j;t)
\approx 
\left(
w+\kappa
\right)^{2(L-j)}
\frac{
e^{-2\Delta_{\rm f/b}t}
t^{2(L-j)}
}
{
4^{L-j}([L-j]!)^2
}
\end{align}

Fixing $L$ and $j$ and differentiating the simplified expression with respect to time and setting it to $0$ yields  Eq.~(11) of the main text, 
\begin{align}\label{eq:crit_t_SM}
    t^{\max}_{L-j}
    \equiv
    \frac{L-j}
    {w\pm \Gamma}.
\end{align}
This expression gives the time at which Eq.~(\ref{eq:Approx_P_Simplified_SM}) peaks. Hence, $t^{\max}_{L-j}$ as the time at which $P(L, j; t)$ maximally contributes to the occupation of site $L$. Since $t^{\max}_{L-j} \propto L - j$, we can interpret the denominator, $w \pm \Gamma$ as a velocity (recall we have set the lattice constant to unity). Note that for perfect non-reciprocity $\kappa \rightarrow w$, this quantity is equal to the dissipative gap $\Delta_{\rm f/b} = \kappa \pm \Gamma \simeq w \pm \Gamma$.

Next, the (maximal) height of $P$ is obtained by plugging Eq.~(\ref{eq:crit_t_SM}) back into Eq.~(\ref{eq:Approx_P_Simplified_SM}). To get a more transparent expression of the height, we write in units of $w = 1$, and take the non-reciprocal limit $\kappa \rightarrow w = 1$. After applying Stirling's approximation to the factorial, we obtain for the height of $P$,
\begin{equation}
\begin{aligned}
P(L, j; t^{\max}_{L-j}) \simeq \frac{1}{2 \pi (L - j)} \frac{1}{\Delta_{\rm f/b}^{2(L - j)}}.
\end{aligned}
\end{equation}
For reasonably large $L - j$, it is clear that the power of $1 / \Delta_{\rm f/b}^{2(L - j)}$ controls whether this expression increases or decreases with $L - j$. Recall that $\Delta_{\rm f/b} = \kappa \pm \Gamma \simeq 1 \pm \Gamma$ in this limit. Hence, for fermions we expect the height to decrease with $L - j$ whereas for bosons we expect the height to increase, as seen in Fig.~\ref{fig:P_fermion_boson}. 

\begin{figure}[htpb]
    \centering
    \includegraphics[width=0.99\linewidth]{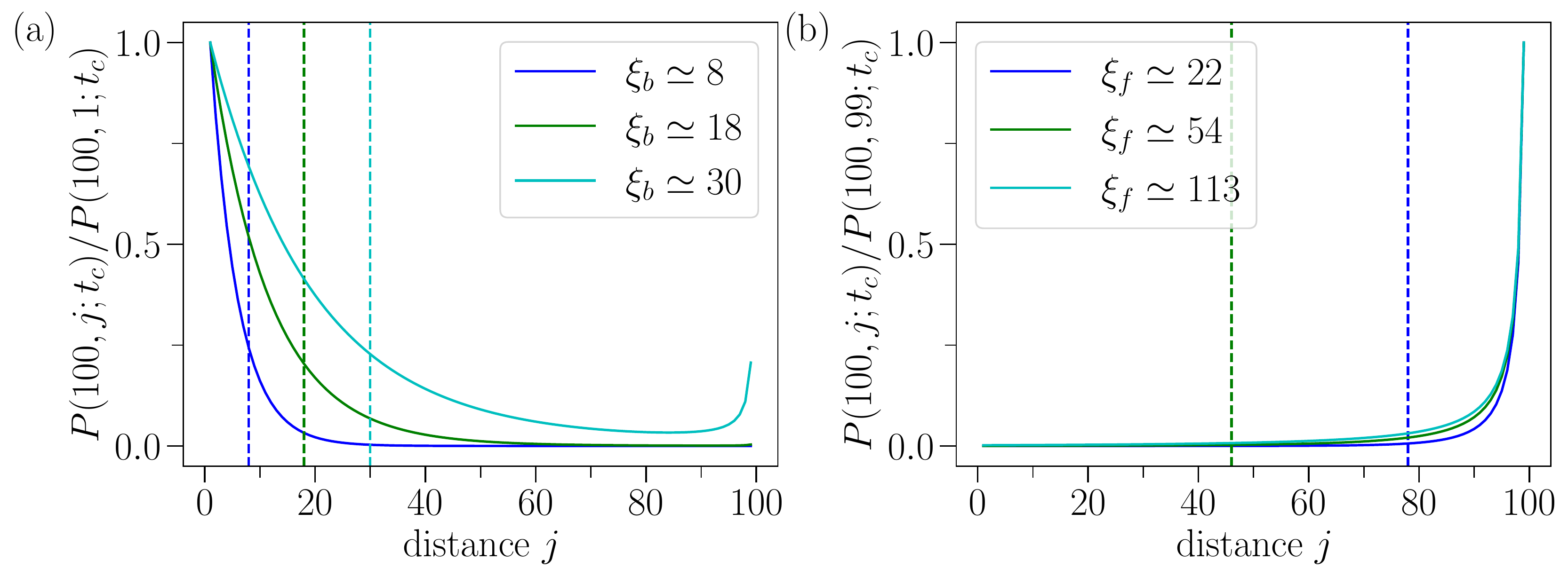}
    \caption{Plot of the peak of the normalized propagators $P(100, j, t_c)/P(100, 99, t_c)$ (see Eq.~(\ref{eq:Approx_P_SM})) describing the propagation of an injected particle from site $j$ to site $100$, against $j$ for bosons (a) and fermions (b). Note that the largest contributions for bosons come from small $j$ (far from site $100$) while the largest contributions for fermions come from large $j$ (close to site $100$). These plots in the highly non-reciprocal limit with $w = 1, \kappa = 0.999$, while the different values of $\xi_{\rm prop}^{\rm b/f}$ are set by varying $\Gamma$. Different parameter regimes are described by their values of $\xi_{\rm prop}^{\rm b/f}$, which describe the characteristic decay length of these peaks. Vertical lines mark $\xi_{\rm prop}^{\rm b/f}$ where visible. }
    \label{fig:height_fermion_boson}
\end{figure}

We can gain further physical insight into the amplification physics by re-writing $1 / \Delta_{\rm f/b}^{2(L - j)}$ as an exponential,
\begin{equation}
\begin{aligned}
P(L, j; t_{c}) \equiv \frac{1}{2 \pi (L - j)} e^{\mp 2(L-j)/\xi_{\rm prop}^{\rm f/b}}.
\end{aligned}
    \label{eqn:height_SM}
\end{equation}
Heuristically, for bosons: when the height of $P$ when $L - j \ll \xi_b$ is exponentially smaller than when $L - j \gg \xi_b$, such that we expect the main contributions to the occupation of site $L$ to come from $L - j \ll \xi_b$ when the chain is long enough. The converse is true for fermions. This is further illustrated in Fig.~\ref{fig:height_fermion_boson}, where we plot the normalized heights $P(100, j; t_{\rm peak})/P(100, 1; t_{\rm peak})$ against $j$ for various values of $\xi_{\rm f/ b}$. Of course, the height is just a crude measure of the contribution, which is in fact an integral over $P$. To gain further insight, we also have to consider the width of $P$.

To estimate the width of $P$, we fit the second derivative of $P$ at $t_{c}$ to a Gaussian, and note that the width of the Gaussian provides an upper bound for the width of $P$, which one can show by comparing derivatives. The second derivative of $P$ at $t_c$ is given by 
\begin{equation}
\partial_t^2 P(L, j; t_c) \simeq - \frac{1}{\pi} \left( \frac{L - j}{\Delta_{f, b}}\right)^{2 (L - j) - 2}
\end{equation}
To approximate this by a Gaussian function of the form $g(t) = A \exp \left( - (t - t_c)^2 / 2 \sigma^2 \right)$, we take $A = P(L - j, t_c)$ and 
\begin{equation}\label{eq:gaussian_sigma_SM}
    \sigma = \sqrt{ - \frac{P(L - j, t_c)}{\partial_t P(L - j, t_c)}} = \frac{\sqrt{L - j}}{\Delta_{\rm f, b}},
\end{equation}
from which we see that the width of $P$ scales as $O(\sqrt{L - j})$. A comparison of the Gaussian approximation against $P$ is plotted in dashed lines in Fig.~\ref{fig:P_fermion_boson}. This $O(\sqrt{L - j})$ behavior is essentially the effect of diffusion. 

\subsubsection{Boson relaxation times}

What is the upshot of describing the propagator $P(L, j; t)$ in such detail? In this section, we focus on the physical interpretation of the propagator for bosons, and argue that these properties allow us to make the approximation that each a particle injected at site $j$ arrives at site $L$ at a single time $t_c$. We refer to this as the single arrival time approximation. 

Since the height of $P(l, j; t)$ for bosons is exponentially larger for $j$ near the left-most part of the chain when $l \gg \xi_{\rm prop}^{\rm b}$, we expect the relaxation times to be controlled by the arrival of particles injected by sites within distance $\xi_{\rm prop}^{\rm b}$ of the start of the chain. This suggests that the relaxation time of the $L$th site on a length $L$ chain scales as $L - \xi_{\rm prop}^{\rm b} \sim  O(L)$. Since the width of $P$ scales as $O(\sqrt{L})$, the arrival of the particle at site $L$ is in fact spread over some time that goes as $O(\sqrt{L})$. This suggests that the single arrival time approximation will get the relaxation time scaling right up to a term that goes as $O(\sqrt{L})$, which will not affect our leading order in $L$ result. 

The single arrival time approximation tells us that the relaxation time of a site $L$ is given by the time it takes a particle to propagate such a distance. From earlier, we know that the speed of this particle in the non-reciprocal limit is $w - \Gamma \simeq \Delta_{\rm b}$. Hence when $L \gg \xi_{\rm prop}^{\rm b}$, we should see relaxation times scaling as Eq.~(\ref{eq:boson_scaling}), 
\begin{align}\label{eq:boson_scaling_SM}
    \tau_{L, \rm b}
    \sim
    t^{\max}_{L-1}
    \approx
    \frac{L}{\Delta_{\rm b}}.
\end{align}

Examining Fig.~1b of the main text, we see that for large $L$, the slope of the relaxation time against chain length always approaches $1/\Delta_{\rm b}$. However, for the plotted values of $L, \Gamma$, there can be a significant displacement of the curve. Furthermore, at small values of $L$, the slope does not appear to be anywhere near $1/\Delta_{\rm b}$.

To explain this behavior, we argue that the relaxation time curve for bosons can be understood as interpolating between two behaviors governed by the relationship between the chain length $L$ and the amplification length-scale $\xi_{\rm b}$. We already understand the $L \gg \xi_{\rm b}$ scaling. To complete the curve, we need to understand the $L \ll \xi_{\rm b} $ scaling. This also allows us to make a concrete statement about how $\xi_b$ controls the relaxation time scales for bosons.

In this regime, we can very roughly approximate the contribution of each site on the chain to the occupation of site $L$ to be proportionate to $1 / \sqrt{L - j}$. To obtain this approximation, we set $e^{2 (L - j) / \xi_{\rm prop}^{\rm b}} \simeq 1$, and note that the area under $P$ can be estimated by its height, which goes as $1 / (L - j)$ multiplied by its width, which goes as $\sqrt{L - j}$, up to some proportionality constant, which we assume to be approximately equal for values of $L - j \ll \xi_{\rm prop}^{\rm b}$.

Then using the estimate $\sum_{k = 1}^n 1 / \sqrt{k} \simeq 2 \sqrt{n}$, we find that the fractional distance from steady state at time $t_{L - j}^{\rm max}$ is approximately $1 - \sqrt{j / n}$. Applying the single arrival time approximation, we find that at time $n (L - j) / \Delta_{\rm b}$, we have
\begin{equation}\label{eq:small_Gamma_approx}
    \delta n_L \left( \frac{L - j}{\Delta_b}\right) \approx 1 - \sqrt{\frac{L - j}{L}}
\end{equation}
To achieve $\delta n_L =  e^{-1}$, we need $j = L (1 - e^{-1})^2$, which gives an estimate of the lower bound of the bosonic slope as $(1 - e^{-1})^2$. The bosonic results from the main text are replotted Fig.~\ref{fig:fermion_sat_fit}a to show that this effectively gives a lower bound on the relaxation time curves for when $\Gamma$ is very small. 

\subsubsection{Fermion relaxation times}

\begin{figure*}
    \centering
    \includegraphics[width=0.75\linewidth]{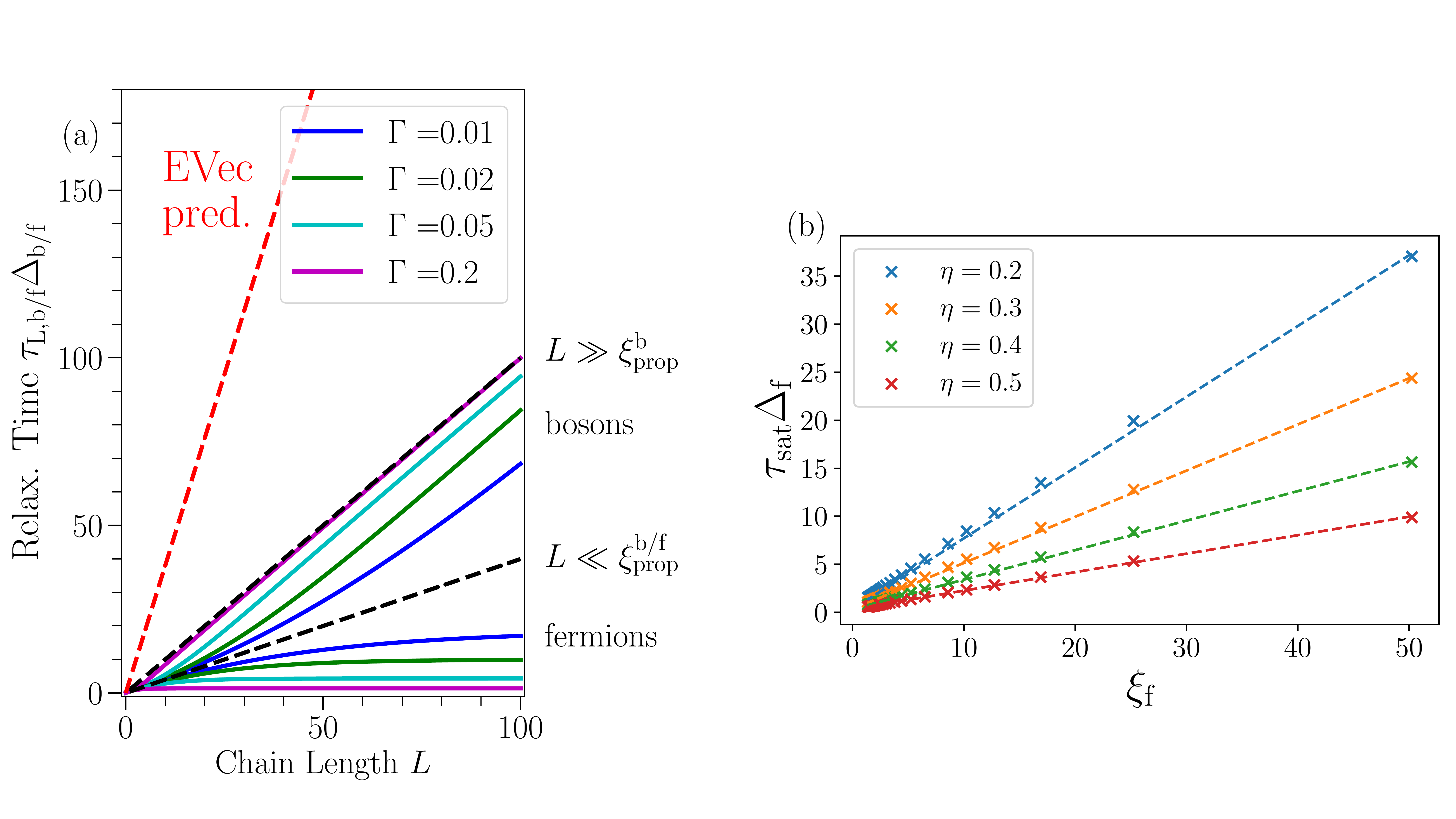}
    \caption{(a) Normalized relaxation time $\tau_{L, {\rm b/f}} \Delta_{\rm b/f}$ for various $\Gamma$ in the highly non-reciprocal limit $w = 1, \kappa = 0.999$ leading to a fixed localization length $\xi_{\rm loc} = 0.26 $. The red dashed line refers to the EVec prediction, the upper black dashed line refers to the scaling for bosons when $L \gg \xi_{\rm prop}^{\rm b}$ and the lower black dashed line refers to the lower (upper) bound for the bosonic (fermionic) curves when $L \ll \xi_{\rm prop}^{\rm b/f}$. We observe that for different values of $\Gamma, L$, the bosonic curves interpolate between the two black dashed lines, and remain strictly between the two, whereas the fermionic curves remain strictly below the lower black dashed line.
    (b) Fermion saturation times $\tau_{\rm sat} \Delta_{\rm f}$ against propagation length $\xi_{\rm f}$. The points in the scatter plot come from numerical calculations, whereas the solid lines are linear fits to the numerics. All $r^2$ values exceed $0.999$, suggesting that in Eq.~\ref{eq:fermion_scaling}, $\tau_{L, \rm f}$ is strongly linear with $t_{\xi_{\rm f}}^{\rm max}$. $\eta$ refers to the relaxation threshold chosen for the numerical calculation. Recall that the relaxation threshold is the value of $\delta n_L$ for which we consider the site to have relaxed. For $\eta = 0.2, 0.3, 0.4, 0.5$, the exact slopes of the linear fits to two decimal places are given by $0.74, 0.48, 0.31, 0.19$.}
    \label{fig:fermion_sat_fit}
\end{figure*}

The only difference between $P$ for bosons and fermions lies in the factor $e^{\pm (L - j)/ \xi_{\rm prop}^{\rm f/ b}}$. Hence, the same approximations we made for bosons when $L \ll \xi_{\rm prop}^{\rm b}$ also apply for fermions when $L \ll \xi_{\rm prop}^{\rm f}$. Hence the above expression Eq.~\ref{eq:small_Gamma_approx} also applies to the fermionic curves when $L \ll \xi_{\rm prop}^{\rm f}$, and the estimated lower bound for the bosonic relaxation curves serves as an upper bound for fermionic relaxation curves. 

We are also interested in the saturation of the fermionic decay times. This occurs when $L \gg \xi_{\rm f}$. As argued in the main text, this results in $\tau_{\rm L, f} \propto \xi_{\rm f} / \Delta_{\rm f}$. This relation is numerically confirmed in Fig.~\ref{fig:fermion_sat_fit}b, where the linear fit is found to satisfy $r^2 > 0.999$ for a range of values of $\Gamma$ and relaxation thresholds $\eta$ for which a saturation time is numerically accessible (i.e. the relaxation time for fermions saturates for a chain of length $L = 180$). The coefficient is monotonically increasing with $1 - \eta$, but its exact value appears to depend on the microscopic details.

\subsubsection{Relaxation curve}

In the main text, we characterized the temporal relaxation of particle density on site $L$ in our model by a single time $\tau_L$.  Here we discuss the full time-dependent relaxation curve.  We show how $\tau_L$ relates to the behaviour of the full curve, and also discuss how our propagator-based approach allows insights into its full form.  

We start by noting that previous studies of relaxation in non-Hermitian models have suggested that such systems often exhibit a two-step relaxation process
(see e.g.~\cite{Song_2019_PRL_chiral_damping, Znidaric_2022_arxiv_randcirc_NHSE}).  In particular, the short-to-intermediate time relaxation is generally characterized by an approximate rate that is completely different than the rate determining the asymptotic long-time relaxation.  It is commonly argued from pseudospectral considerations that the asymptotically long relaxation rate is given by the dissipative gap \cite{trefethen_pseudospectra_2005}.  The relaxation metric $\tau_L$ we study in the main text (corresponding to a $1/e$) decay is associated with short-to-intermediate time behaviour, and not the asymptotic long-time decay rate.  For the models we study, the long-time relaxation behaviour (controlled by the dissipative) gap only emerges when one is exponentially close to the the steady state, making it of little interest (and even difficult to numerically extract for general parameters). 

\begin{figure*}
    \centering
    \includegraphics[width=0.75\linewidth]{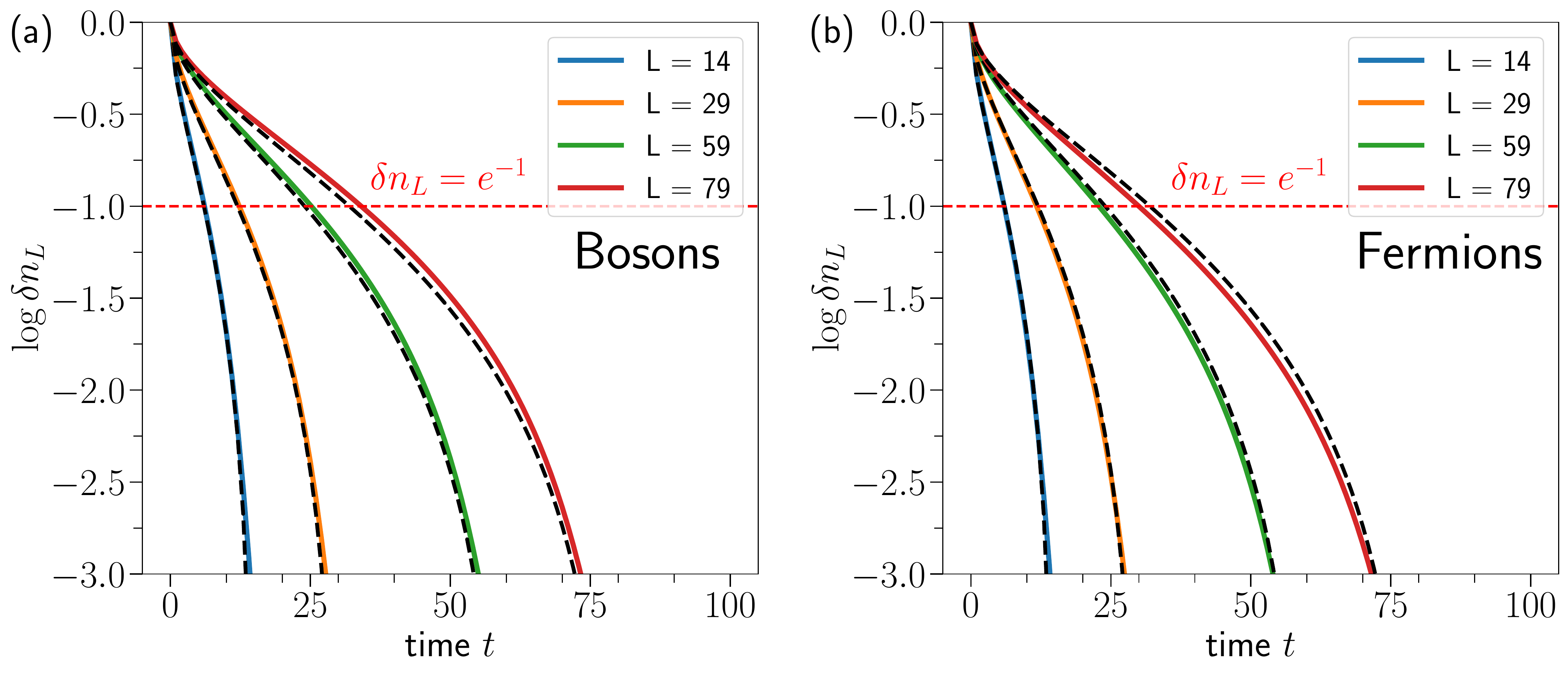}
    \caption{(a) $\log \delta n_L$ against time for bosons in the in the highly non-reciprocal limit $w = 1, \kappa = 0.999$ with small pumping $\Gamma = 0.001$, where $\delta n_L$ is the normalized deviation of site $L$ from its steady state value. The solid coloured lines indicate numerically simulated results whereas the black dashed lines indicate the estimate given in Eq.~\ref{eq:small_Gamma_est}. The red dashed line indicates our chosen relaxation threshold of $e^{-1}$. For bosons, the numerical curves lie slightly above the estimated curve, and approach it as $\Gamma \rightarrow 0$. (b) Same as (a) but for fermions. For fermions, the numerical curves lie slightly below the estimated curve, and also approach it as $\Gamma \rightarrow 0$.  }
    \label{fig:SM_time_dep_1}
\end{figure*}

Given this, let us consider how our approach allows us to understand the short to intermediate time relaxation behavior. For small $\Gamma$ (i.e. almost no pumping), Eq.~(\ref{eq:small_Gamma_approx}) allows us to approximate the relaxation curve up to the relaxation threshold $\tau_L$ and beyond. Substituting $t = (L - j) / \Delta_{\rm f/b} $ into Eq.~(\ref{eq:small_Gamma_approx}), we obtain the estimate 
\begin{equation}\label{eq:small_Gamma_est}
\delta n_L(t) \simeq 1 - \sqrt{\frac{\Delta_{\rm f/b}t}{L}}.
\end{equation}
The accuracy of this estimate is illustrated in Fig.~\ref{fig:SM_time_dep_1} for both bosons and fermions, where it is compared against the full, numerically-simulated relaxation curve for various sites at a small $\Gamma = 0.001$ value. We see that Eq.~(\ref{eq:small_Gamma_est}) gives us an excellent approximation for such small values of $\Gamma$ up to intermediate timescales, becoming virtually exact when $\Gamma \rightarrow 0$. The relaxation time we have identified can then be read off from Eq.~(\ref{eq:small_Gamma_est}), and offers a coarse-grained means to exactly understand the zero pumping limit.

\begin{figure*}
    \centering
    \includegraphics[width=0.75\linewidth]{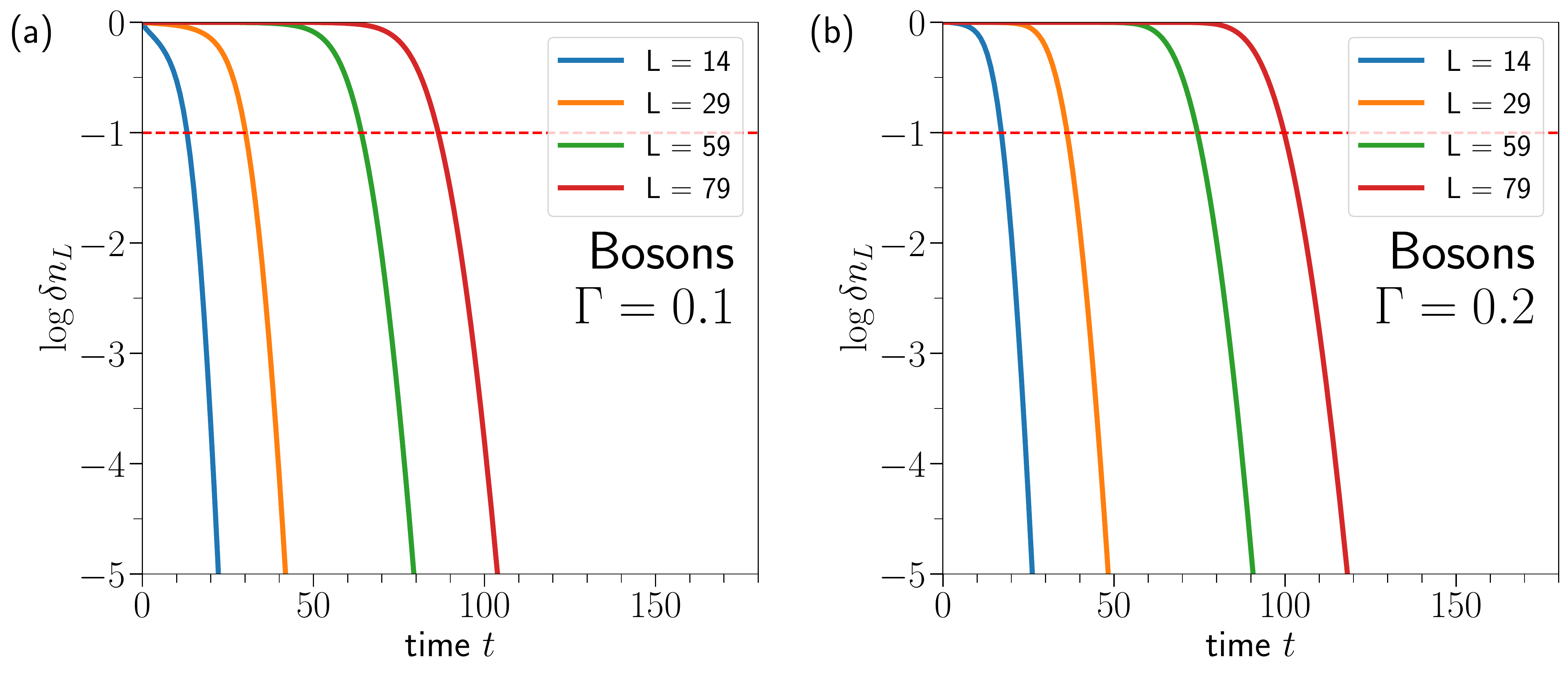}
    \caption{(a) $\log \delta n_L$ against time for bosons in the highly non-reciprocal limit $w = 1, \kappa = 0.999$ when $\Gamma = 0.1$, where $\delta n_L$ is the normalized deviation of site $L$ from its steady state value. The solid coloured lines indicate numerically simulated results and the red dashed line indicates our chosen relaxation threshold of $e^{-1}$. We observe that the relaxation curves start out with essentially flat before suddenly shooting past the relaxation threshold, which corresponds to the timescale at which we begin to see the effects of amplification. (b) Same as (a) but for $\Gamma = 0.2$.}
    \label{fig:SM_time_dep_2}
\end{figure*}

For large $\Gamma$, where the length of the chain $L$ is comparable to or less than the propagation lengthscale $\xi_{\rm prop}^{\rm b}$, the initial slope of relaxation curve for bosons approaches $0$ as site number increases, before sharply increasing in magnitude.  This transition in slope almost exactly coincides with our relaxation time $\tau_L$, as can be seen in Fig.~\ref{fig:SM_time_dep_2}. Our propagation-based picture explains the initial flat slope as the period before exponential amplification kicks in:
during this short-time period, the amplified contribution of particles injected near the left edge of the system has not yet reached the site of interest.  As argued in the main text, the length-scale $\xi^{\rm prop}$ that we have identified helps us identify the transition between these two regions of decay. In this picture, the relaxation time $\tau_L$ identifies the timescale at which we begin to observe the effects of exponential amplification.

\begin{figure*}
    \centering
    \includegraphics[width=0.75\linewidth]{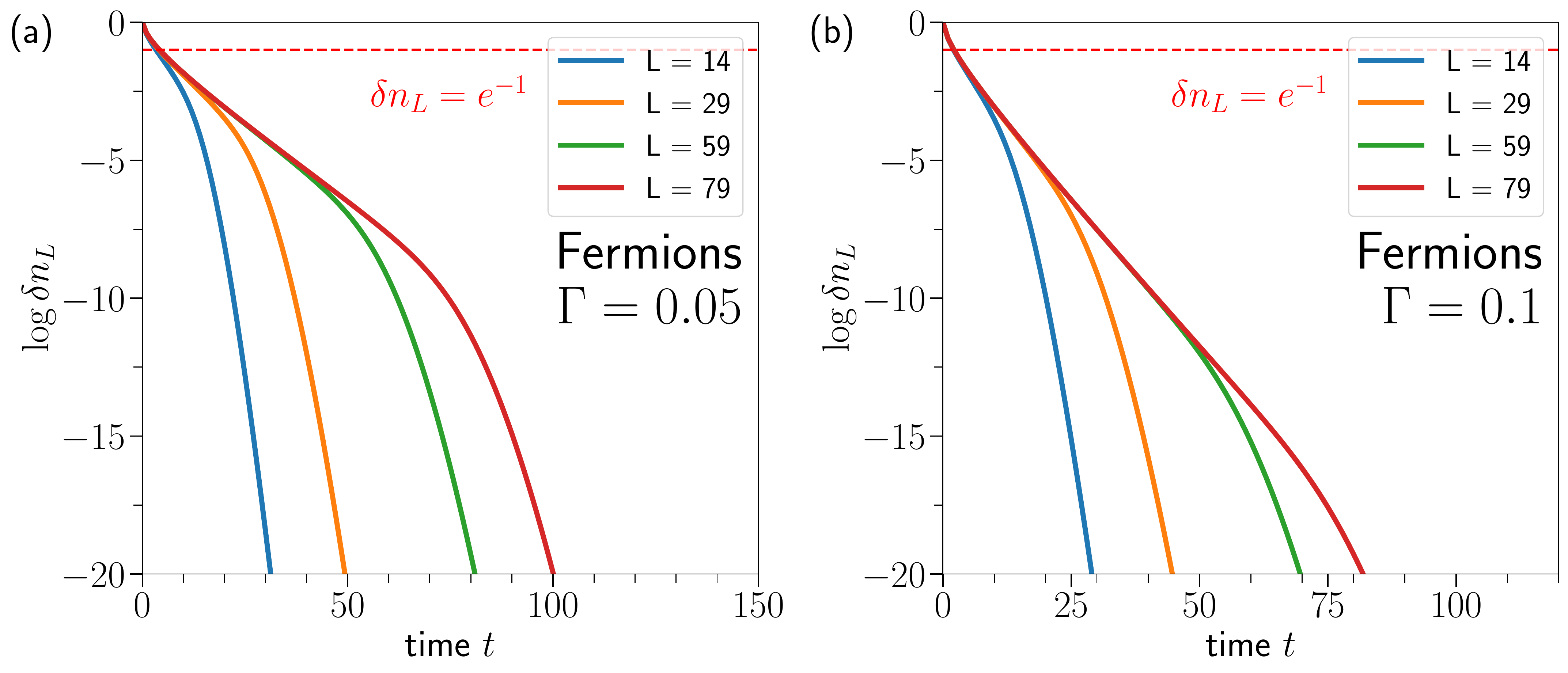}
    \caption{(a) $\log \delta n_L$ against time for fermions in the highly non-reciprocal limit $w = 1, \kappa = 0.999$ when $\Gamma = 0.05$, where $\delta n_L$ is the normalized deviation of site $L$ from its steady state value. The solid coloured lines indicate numerically simulated results and the red dashed line indicates our chosen relaxation threshold of $e^{-1}$. We observe that the relaxation across a large range of sites start out with a similar curve, and cross the relaxation threshold at the same time, corresponding to our observation that the relaxation time for fermions plateaus with $L$. One can observe two different regimes of relaxation for fermions characterized by different exponential rates. (b) Same as (a) but for $\Gamma = 0.1$.}
    \label{fig:SM_time_dep_4}
\end{figure*}

Finally, for fermions, Ref.~\cite{Song_2019_PRL_chiral_damping} previously observed a two-step relaxation in a model that is closely related to our quantum Hatano Nelson chain (namely a non-reciprocal dissipative SSH chain).  Crucially, their study focused on the regime of extremely weak non-reciprocity, whereas we are most interested in regimes of strong non-reciprocity.  
Ref.~\cite{Song_2019_PRL_chiral_damping} argues that at short times, and in this weak non-reciprocity regime, exponential temporal decay of local amplitudes can be compensated for exactly by the effects non-reciprocal propagation, resulting in a power law decay. For strong non-reciprocity, such a complete cancellation is no longer possible.  As a result, even at short times the relaxation curve can exhibit exponential decay. However, at short times, there can still be a partial compensation, and as seen in Fig.~\ref{fig:SM_time_dep_4}, fermion relaxation trajectories are typically characterized by two different exponential rates in the short and intermediate times. In this picture, again our relaxation time scale corresponds to the short to intermediate time relaxation curve which is almost identical for all sites at different chain lengths when $\Gamma$ is large. Alternatively, we can say that our relaxation time scale characterizes the first step of this two-step relaxation process, as is evident from where the dashed red line, indicating $\delta n_L = e^{-1}$, crosses the relaxation curves in Fig.~\ref{fig:SM_time_dep_4}.

\section{Other initial conditions} \label{app:other_initial_conditions}

\begin{figure}
    \centering
    \includegraphics[width=0.99\linewidth]{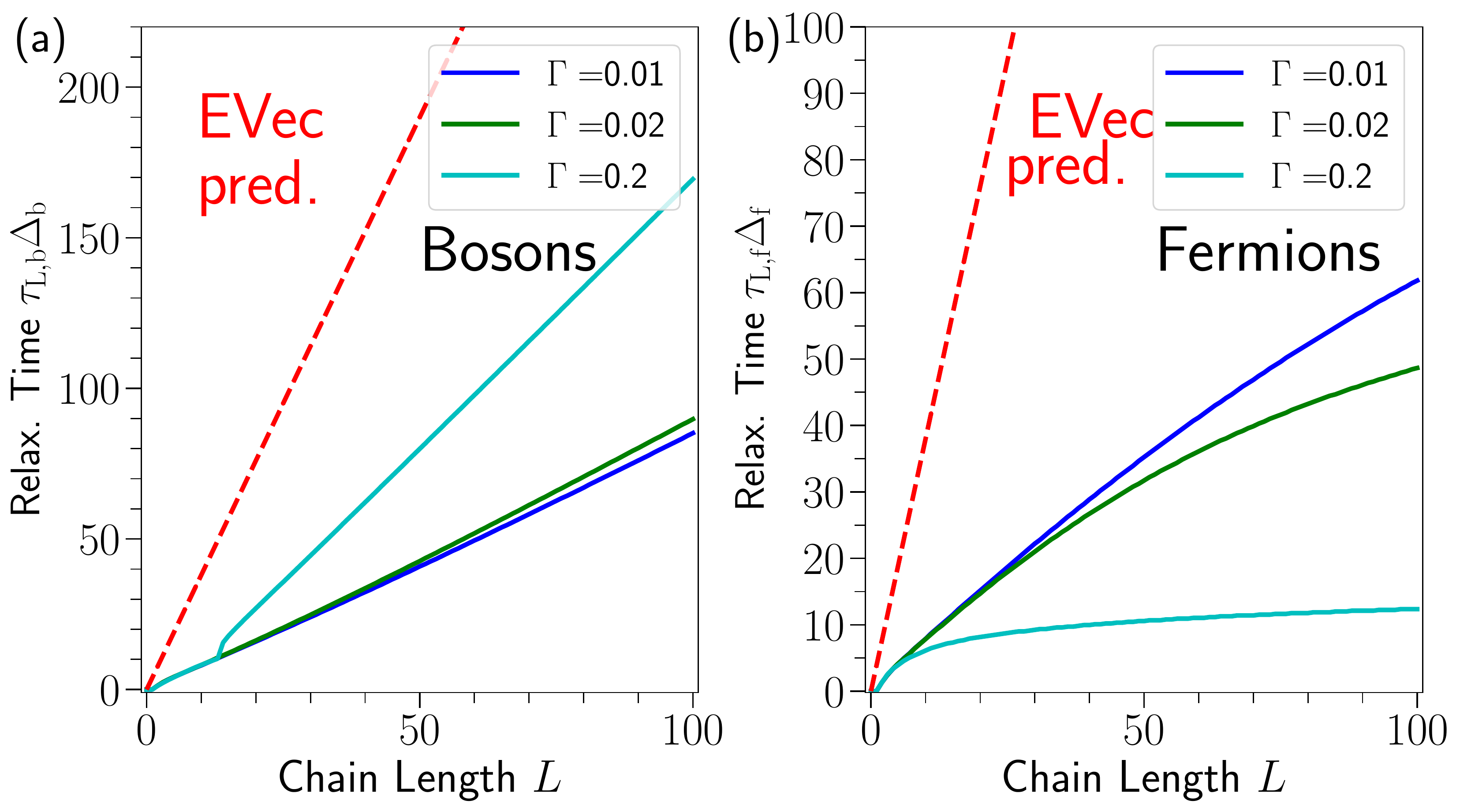}
    \caption{Normalized relaxation time $\tau_{L, {\rm f/b}}$ for various $\Gamma$ in the highly non-reciprocal limit $w = 1, \kappa = 0.999$, for initial state being a uniform distribution given by the average occupation of the steady state, for (a) bosons, (b) fermions.}
    \label{fig:alt_i_1}
\end{figure}

\begin{figure}
    \centering
    \includegraphics[width=0.99\linewidth]{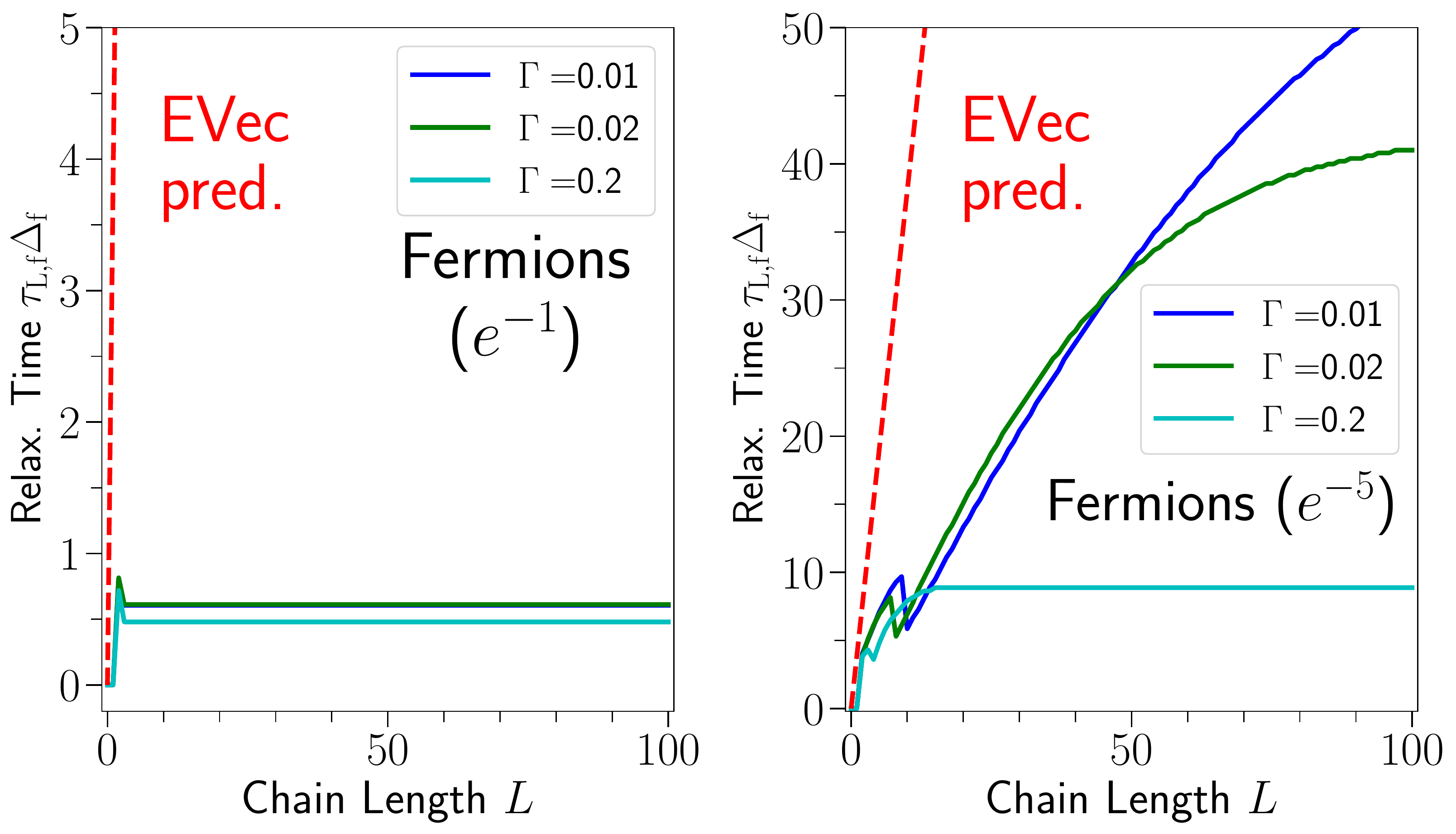}
    \caption{Normalized relaxation time $\tau_{L, {\rm f}}$ for various $\Gamma$ in the highly non-reciprocal limit $w = 1, \kappa = 0.999$, for initial state being the all-filled state for (a) relaxation threshold of $e^{-1}$, (b) relaxation threshold of $e^{-5}$.}
    \label{fig:alt_i_2}
\end{figure}

In this section, we present numerical relaxation times in our model for two other natural initial conditions which are distinct from the vacuum. First, we consider the case where the initial condition comprises a uniform distribution with average density equal the steady state average density, that is, $\langle \hat{c}_n^{\dag} (0) \hat{c}_m(0) \rangle = \delta_{nm} \bar{n}_{\rm ss}(L)$, where $\bar{n}_{\rm ss}(L) = \frac{1}{L} \sum_{i = 1}^{L} \langle \hat{c}_i^{\dag} \hat{c}_i \rangle_{\rm ss}$. Second, for fermions we also consider the case where the chain starts out in the all-filled state. 

The normalized relaxation times as a function of chain length for the initial condition $\langle \hat{c}_n^{\dag} (0) \hat{c}_m(0) \rangle = \delta_{nm} \bar{n}_{\rm ss}(L)$ are plotted in Fig.~\ref{fig:alt_i_1}. We note that for these initial conditions, the relaxation of the chain may not be monotonic. This is because there are two mechanisms which lead to a change in on-site occupation: one contribution comes from the initial distribution of particles, and another comes from the pump baths adding particles to each site. To see this concretely, consider formally solving  Eq.~(3) of the main text. Let $\mathbb{S}(t)_{nm} \equiv \langle \hat{c}_n^{\dag}(t) \hat{c}_m (t) \rangle $ be the normal-ordered correlation matrix. Then the solution can be written as,
\begin{widetext}
\begin{equation}\label{eq:formal_eom_soln_0}
\begin{aligned}
    \mathbb{S}(t) = 
    & \exp(- i \mathbb{H}_{\rm HN} t ) \mathbb{S}(0) \exp( i \mathbb{H}_{\rm HN} ^{\dag} t) 
    + \int\limits_0^t dt' \exp(- i \mathbb{H}_{\rm HN} (t - t')) 2 \Gamma \exp(i \mathbb{H}_{\rm HN}^{\dag}(t - t')),
\end{aligned}
\end{equation}
\end{widetext}
where the first term containing $\mathbb{S}(0)$ corresponds to the contribution coming from the initial state and the second term containing $\Gamma$ corresponds to the contribution from the pump. The matrix element $(\mathbb{S}(t))_{L, L}$ gives the occupation of the $L$-th site at time $t$.  We focus on the contributions of the initial state term and the pumping term to this quantity.

From our study of the chain initialized from vacuum, we know that the pumping term contributes to the occupation in a monotonic fashion . We thus focus on the behavior of the initial state term, which is what gives the non-monotonic behavior. For bosons, in the limit of large non-reciprocity, occupation on the last site is much larger than on all other sites and thus $\overline{n}_{\rm ss} \lessapprox \langle \hat{n}_L \rangle_{\rm ss} $. Hence,  as the particles from the initial state are redistributed in the lattice across the chain, we find that we very quickly reach and then exceed the steady-state occupation on the last site, only to once again relax towards the steady state. Our chosen metric thus does not accurately capture the relaxation dynamics. 

To numerically account for this potential non-monotonic behavior, we further impose the condition that the site remain relaxed for some amount of time after first achieving the relaxation condition. In particular, for our simulations, we fix this to be three times the time at which the relaxation condition is first met, i.e. if the occupation of the site is close enough to the steady state that it appears to be relaxed at some time $t$, we require that it stays relaxed for a time $3t$ before counting $t$ as the relaxation time. For bosons under the first kind of initial conditions, for numerical tractability we only simulate relaxation times for a maximum of $\Gamma = 0.1$. Due to the interplay of the two terms in Eq.~(\ref{eq:formal_eom_soln_0}), we find that this initial condition generally leads to longer relaxation times compared to starting from the vacuum. 

The normalized relaxation times against chain length for the second type of initial condition are plotted in Fig.~\ref{fig:alt_i_2}. For the second type of initial condition, we note that fermions relax to the steady state extremely fast. This can be understood as the combination of two effects. First, all sources of dissipation $\Gamma, \kappa, \lambda$ contribute to loss. If we start the chain in the all-filled state, the site occupation of the fermions monotonically approaches the steady state from above. Second, for high non-reciprocity, the steady state occupation of the last site approaches $1$ quickly as $L$ increases, even for relatively small values of $\Gamma$ \cite{AM_QHN_paper}. Hence, the steady state is easier to achieve when starting out from the all-filled state. In Fig.~\ref{fig:alt_i_2}a, we use the same $e^{-1}$ relaxation threshold as the main text. To resolve the relaxation behavior in more detail, in Fig.~\ref{fig:alt_i_2}b, we use a much more stringent $e^{-5}$ relaxation threshold.

For these different initial conditions, we note that the same qualitative behavior -- linear scaling for the bosonic relaxation curves and eventual saturation of the fermionic relaxation curves -- holds as in the main text, demonstrating that our claims are independent of the specific initial conditions.

\section{Other models}\label{app:other_models}

The quantum Hatano-Nelson model is special in that, under open boundary conditions, we can identify a local non-unitary transformation which maps the Hatano-Nelson model to a trivial reciprocal Hermitian tight-binding model. In the process, this yields an analytical expression for a localization length $\xi_{\rm loc}$. 
Unfortunately, this is not possible in a general non-Hermitian model. One can still identify a localization length numerically. 

In this section, we will examine two additional models. The first is the non-Hermitian SSH model, for which a similar local non-unitary transformation is possible. One key qualitative distinction of the non-Hermitian SSH model from the Hatano-Nelson model is that it has two bands.  The second is the next-nearest neighbor Hatano-Nelson model, for which we have to numerically extract the localization length. For a quantum mechanically consistent effective Hamiltonian given by the two models, we follow the minimal prescription given in \cite{AM_QHN_paper}.

The goal of this section is to show that the EVec prediction generically does not accurately predict relaxation times. We will do so by numerically calculating the same analogous quantities of site relaxation time and localization length for each of these two models. In general, for strong NHSE, the EVec prediction is an overestimate. Furthermore, similar to the Hatano-Nelson model in the main text, we find that by tuning local loss and gain parameters, one can vary the observed relaxation times without changing the eigenvectors of the system.

\subsubsection{Non-Hermitian SSH model}

The effective Hamiltonian for the non-Hermitian SSH model \cite{lieu2018_nh6, AM_2022_unpublished, Yang_2022_PRR} is 
\begin{widetext}
\begin{equation}
    \begin{aligned}
    \mathbb{H}_{\rm SSH} = &\sum_j \left( \frac{w - \kappa}{2} | A, j \rangle \langle B, j | + \frac{w + \kappa}{2} | A, j \rangle \langle  B, j | \right) 
    + \sum_j \left( \frac{u - \gamma}{2} | A, j  + 1 \rangle \langle  B, j | + \frac{u + \gamma}{2} | A, j + 1 \rangle \langle  B, j | \right) 
    \\&-i \sum_j \left( \pm \Gamma + \frac{\kappa + \gamma}{2} \right) ( | A, j \rangle \langle  A, j | + | B, j \rangle \langle  B, j | ),
    \end{aligned}
\end{equation}
\end{widetext}
where $A, B$ label the two distinct sites in the unit cell, $w, u$ are the bare hoppings and $\kappa, \gamma$ their non-Hermitian modifications, with the $+ ( - )$ sign holding for fermions (bosons). The quantity of interest is the longest local relaxation time for a non-reciprocal SSH chain of length $L$. This Hamiltonian should be understood as an effective Hamiltonian governing the evolution of the covariance matrix, in exact analogy to Eq.~(4) of the main text.

Following \cite{Yang_2022_PRR}, there are two exponentially localized boundary modes, for which we can identify the localization lengths relevant to the EVec prediction via 
\begin{equation}
    \begin{aligned}
    &e^{1/\xi_1} = \frac{u + \gamma}{w - \kappa} \qquad
    && e^{1/\xi_2} = \frac{w + \kappa}{u - \gamma} 
    \end{aligned}
\end{equation}

The EVec prediction goes as
\begin{equation}
    \tau \sim \frac{L}{\Delta \cdot \mbox{min}(\xi_1, \xi_2) }
    \label{eq:LSE_SSH}
\end{equation}
where the dissipative gap is $\Delta = \pm \Gamma + \frac{\kappa + \gamma}{2}$. 

\begin{figure}
    \centering
    \includegraphics[width=0.99\linewidth]{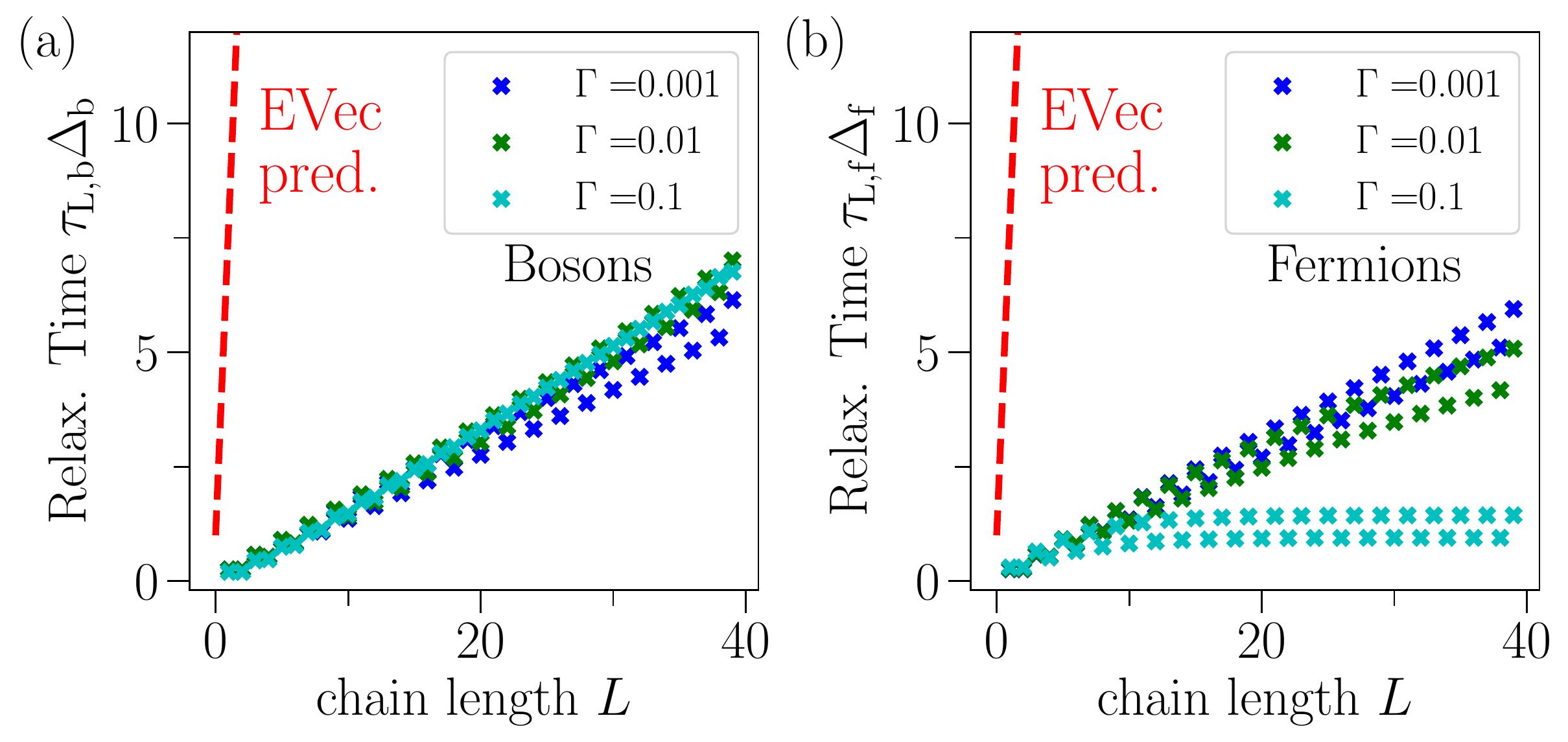}
    \caption{(a) The longest decay time $\tau_{\rm L, b}$ for bosons on a non-reciprocal SSH chain of length $L$ in units of dissipative gap $\Delta_{\rm b}$, with parameters given by $w = u = 1$, $\kappa = 0, \gamma = 0.999$ for various values of $\Gamma$. Numerically obtained values are given by the scatter plots whereas the EVec prediction is given by the dashed lines. Since $\Gamma$ does not affect the localization length, the EVec prediction collapses into the same line for all three values of $\Gamma$. In all cases, the EVec prediction is a severe overestimate, although it gets the linear in $L$ scaling right. (b) Same as (a) but for fermions. We observe that for larger values of $\Gamma$, the relaxation times saturate, as with the quantum Hatano-Nelson chain. }
    \label{fig:nr_SSH_combined_plots}
\end{figure}

In Fig~\ref{fig:nr_SSH_combined_plots}, we plot the longest decay times in units of  the dissipative gap for a range of chain lengths for fermions and bosons, with $w = u$, $\kappa = 0$, and $\gamma = 0.999$, restricting the non-reciprocal hopping to only every other bond, for a range of values of pumping $\Gamma$. In all cases we see that Eq.~(\ref{eq:LSE_SSH}) severely overestimates decay times, and fails to capture the saturation of the relaxation times for fermions, similar to the overestimate for the QHN model in the main text.

\begin{figure}[htpb]
    \centering
    \includegraphics[width=0.99\linewidth]{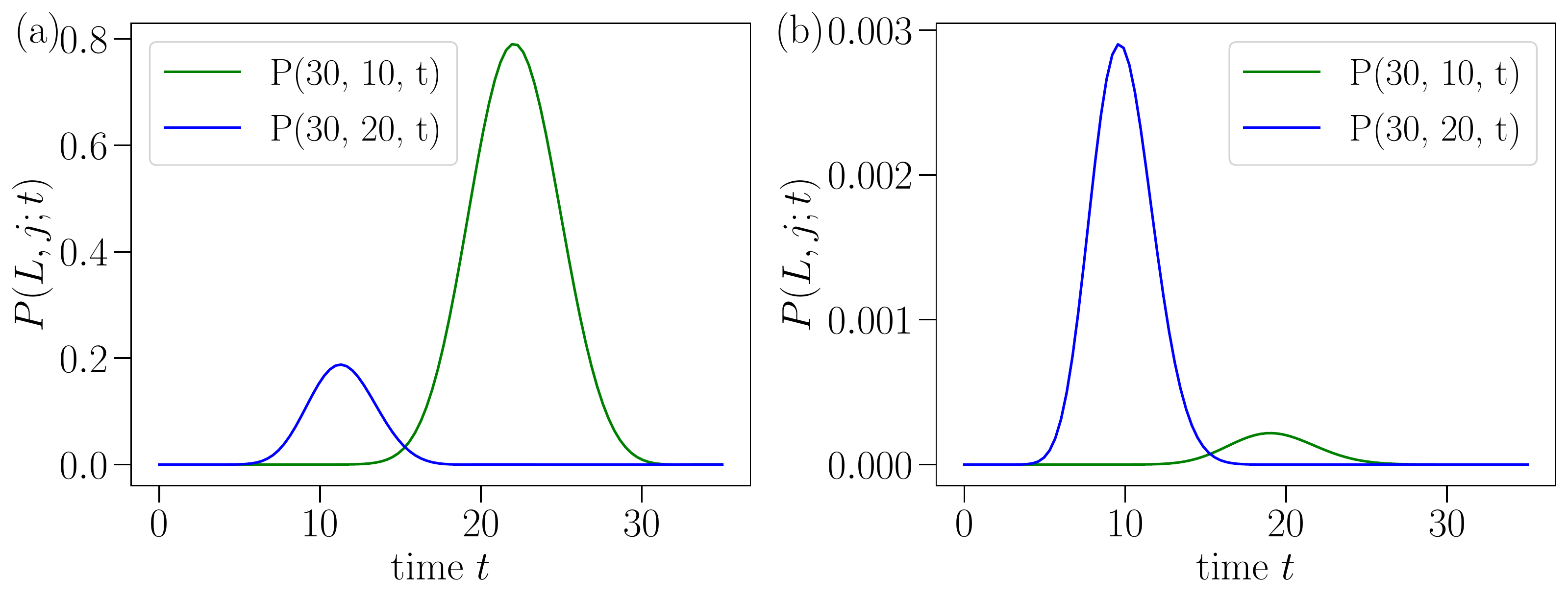}
    \caption{Plot of the time-dependence of the propagators $P(30, 10; t)$ and $P(30, 20; t)$ in the non-Hermitian SSH model for bosons (a) and fermions (b) for parameters $L = 30, w = u = 1, \kappa = 0.5, \gamma = 0.99, \Gamma = 0.1$. Notice that the essential features of the QHN propagators are retained. Similar to the propagators for the QHN model, the functions have a single large peak, and the area under the curve is much larger for $P(30, 10; t)$ than $P(30, 20; t)$ for bosons, whereas the opposite is true for fermions, reflecting the amplification dynamics of the non-Hermitian SSH model. }
    \label{fig:P_fermion_boson_SSH}
\end{figure}

\begin{figure}[htpb]
    \centering
    \includegraphics[width=0.99\linewidth]{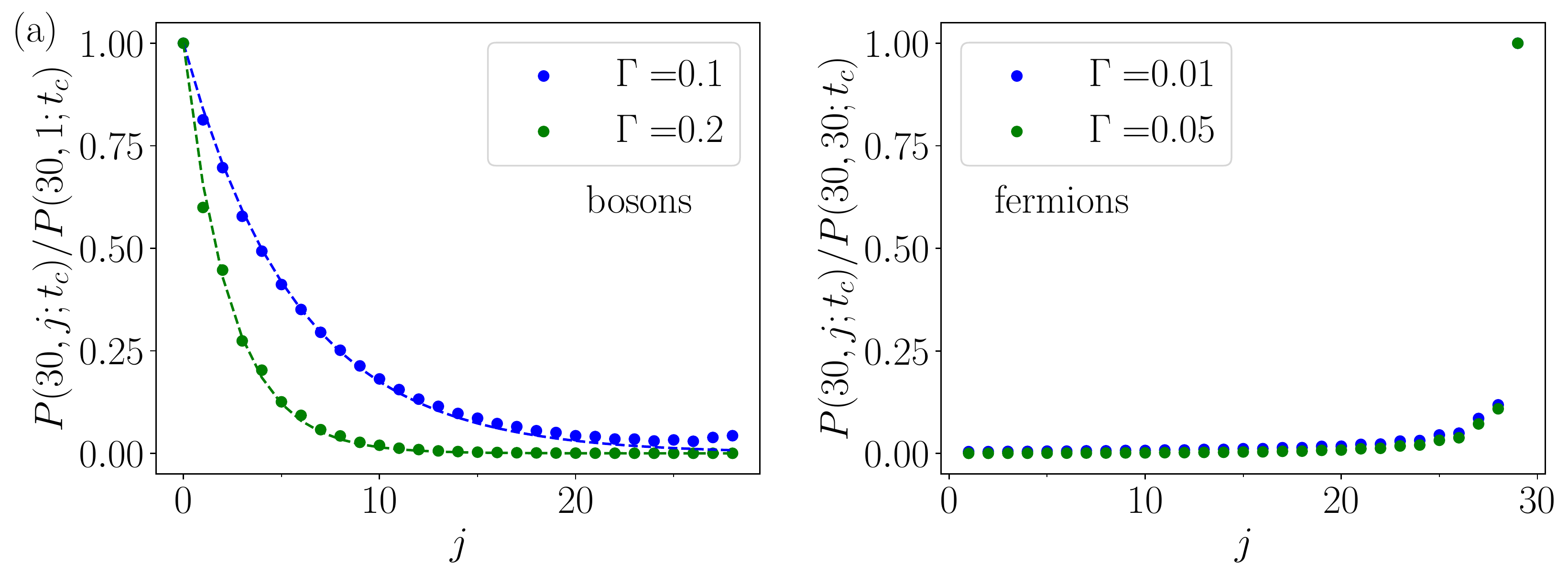}
    \caption{Plot of the peak of the normalized propagators ($P(30, j, t_c)/P(30, 99, t_c)$ for bosons, $P(30, j, t_c)/P(30, 30, t_c)$ for fermions) describing the propagation of an injected particle from site $j$ to site $30$, against $j$ for bosons (a) and fermions (b). As with the QHN model, the largest contributions for bosons come from small $j$ (far from site $30$) while the largest contributions for fermions come from large $j$ (close to site $30$). The parameters used are $L = 30, w = u = 1, \kappa = 0.5, \gamma = 0.99$, with $\Gamma$ indicated in the legend. Notice that $\Gamma$ controls the shape of the plots. For bosons, we numerically fit an exponential decay curve (dashed lines in (a)) to the left half of the chain ($j = 1$ to $j = 15$), and estimate the propagation lengths $\xi_{\rm prop}^{\rm (est)} \simeq 5.75$ corresponding to $\Gamma = 0.1$, and $\xi_{\rm prop}^{\rm (est)} \simeq 2.37$ corresponding to $\Gamma = 0.2$.}
    \label{fig:height_fermion_boson_SSH}
\end{figure}

Next, we will demonstrate that this discrepancy in relaxation times can be understood from the same physics that governs relaxation times in the QHN model. To that end, we carry out numerical investigation of the Green's functions of the SSH model. The quantity of interest here is once again the propagator $P(L, j; t) \equiv | \langle m | e^{ - i \mathbb{H}_{\rm eff} t} | j \rangle |^2$.

For concreteness, we fix $L = 30$, $w = u = 1$, $\gamma = 0.99$, $\kappa = 0.5$. Note that numerically computing the propagators requires us to expand quantities in terms of the eigenvectors of $\mathbb{H}_{\rm eff}$. Since the exponential factors $\sim e^{L / \xi_{1, 2}}$ quickly get very large, we have to keep $L$ relatively small. On a similar note, we will require a propagation length scale that is smaller than $L$, for which we have selected an appropriate $\Gamma$. First, in analogy with Fig.~\ref{fig:P_fermion_boson}, we plot $P(30, 10, t), P(30, 20, t)$ against time for both bosons and fermions in the SSH model in Fig.~\ref{fig:P_fermion_boson_SSH}. We observe that the essential features of these propagators are the same as the QHN propagators, and the relative size of the contributions to the right-most site occupation reflects the amplification dynamics of the non-reciprocal chain. 

Next, in analogy with Fig.~\ref{fig:height_fermion_boson}, we also plot the normalized maximum heights of the propagators against the starting point $j$ in Fig.~\ref{fig:height_fermion_boson_SSH}. As expected, the largest contributions to the right-most site for bosons come from the sites furthest to the left, whereas the contributions for fermions quickly die off with distance. Further, one can easily obtain a numerical estimate of the propagation length for bosons. For sites on the left half of the chain, we expect the maximum heights of the propagators to be dominated by the exponential amplification dynamics. Hence, we can fit an exponential curve $e^{-(L-j)/ \xi_{\rm prop}^{\rm 
(est)}}$ to the heights of the propagators associated with the left half of the chain, as depicted by the dashed lines in Fig.~\ref{fig:height_fermion_boson_SSH}a. This yields the estimates $\xi_{\rm prop}^{\rm (est)} \simeq 5.75$ corresponding to $\Gamma = 0.1$, and $\xi_{\rm prop}^{\rm (est)} \simeq 2.37$ corresponding to $\Gamma = 0.2$. Note that these length scales are small because we have chosen relatively large $\Gamma$ to ensure that the amplification dynamics show up on the scale of the relatively small chain length $L = 30$. However, even at this level we can observe that the length scale can change drastically even though $\xi_{\rm loc}$ is unaffected by $\Gamma$. For fermions, the same numerical estimate is not as easily obtained, since the contributions are doubly damped from both the non-reciprocal dynamics and the underlying reciprocal tight-binding dynamics. 

\subsubsection{Next-nearest neighbor Hatano-Nelson model}

The effective Hamiltonian for the next-nearest neighbor Hatano-Nelson model \cite{AM_QHN_paper} is given by
\begin{widetext}
\begin{equation}
    \begin{aligned}
    \hat{H}_{\rm eff} =
    & \sum_j \left( \frac{w + \kappa}{2} |j + 1 \rangle \langle j | + \frac{w - \kappa}{2} |j \rangle \langle j + 1 | \right) 
    + \frac{T}{2} 
    \sum_j \left( e^{i \phi}|j + 2 \rangle \langle j |+ e^{-i\phi} |j \rangle \langle j + 2 | \right) 
    - i (\pm \Gamma + \kappa) \sum_j |j \rangle \langle j |
    \end{aligned}
\end{equation}
\end{widetext}
where $T$ gives the hopping amplitude for the next-nearest neighbors, and $\phi$ the phase of that hopping. For $\phi \neq 0, \pi$, the phase cannot be gauged away. As established in \cite{AM_QHN_paper}, this can lead to exotic consequences in the steady state. For instance, when $\phi = \pi/2$, then despite non-reciprocity heavily favoring the nearest-neighbor hoppings to the right, we find in the steady state that particles localize to the left.

In this work, we are concerned with the scaling of the longest relaxation times of a given chain length and how they relate (or not) to the localization length. In the next-nearest neighbor QHN model, we cannot perform a gauge transformation that conveniently identifies a localization length. Instead, we have to do so numerically. Consider a 1D effective non-Hermitian Hamiltonian $H_{\rm eff}$, with eigenenergies and right eigenvectors in real space given by $E_{\alpha}, | \psi_{\alpha} \rangle$, with real-space wavefunction $\psi_n^{(\alpha)} \equiv \langle n | \psi_{\alpha} \rangle$. One can write $\psi_n^{(\alpha)}$ as a linear combination of plane waves with complex exponents,
\begin{equation}\label{eq:psi_ansatz}
    \psi_n^{(\alpha)} \propto \sum_{i = 1}^{d} e^{- n A_i^{(\alpha)}(E) + i n B_i^{(\alpha)}(E)},
\end{equation}
where the number of terms $d$ in the sum is finite and depends on the model and $A_i, B_i$ are real numbers, and one can analytically show that each eigenvector is a finite sum of these complex plane waves. Using  $H_{\rm eff} | \psi \rangle = E | \psi \rangle$, one can write an expression involving $A_i, E$, which can be solved numerically for $A_i(E)$. This can be identified with the localization length of $\psi$. This yields a procedure to numerically extract the localization length of any such non-reciprocal model. 

To illustrate and benchmark the method, we start with the Hatano-Nelson model with OBC, for which the localization lengths are analytically known. Recall the effective Hatano-Nelson Hamiltonian,
\begin{widetext}
\begin{equation}
\hat{H}_{\rm HN} = \sum_j \left( \frac{w + \kappa}{2}  | j + 1 \rangle \langle j | + \frac{w - \kappa}{2} |j \rangle \langle j + 1 | - i ( \kappa + \lambda - \Gamma ) | j \rangle \langle j | \right).
\end{equation}
\end{widetext}

Let $| \psi \rangle$ be an eigenvector with components $\psi_j \equiv \langle j | \psi \rangle$. The eigenvalue equation is 
\begin{widetext}
\begin{align}
  \sum_j E  \psi_j | j \rangle &= \sum_j \left(  \frac{w + \kappa}{2}  \psi_j | j + 1 \rangle + \frac{w - \kappa}{2} \psi_{j + 1} |j \rangle - i ( \kappa + \lambda - \Gamma )  \psi_j | j \rangle \langle j |  \right) \\
  (E + i (\kappa + \lambda - \Gamma)) \psi_j &= \frac{w + \kappa}{2} \psi_{j - 1}  + \frac{w - \kappa}{2} \psi_{j + 1} 
\end{align}
\end{widetext}

We use the ansatz $\psi_j \sim c \lambda^j$. The above becomes
\begin{equation}
   0 = \frac{w - \kappa}{2} \lambda^2 - (E + i (\kappa + \lambda - \Gamma)) \lambda +  \frac{w + \kappa}{2}  
  \label{eqn:QHN-quad}
\end{equation}
which has two independent solutions $\lambda_{1, 2}$, so that we can write 
\begin{equation}
\psi_j = c_1 \lambda_1^j + c_2 \lambda_2^j.
\end{equation}

We have two boundary conditions, $\psi_0 = \psi_{N + 1} = 0$. The first gives 
\begin{equation}
c_1 = - c_2 \equiv c
\end{equation}
Then using $\psi_j = c ( \lambda_1^j - \lambda_2^j )$, the second boundary condition gives 
\begin{equation}
\lambda_1^{N + 1} = \lambda_2^{N + 1}
\end{equation}

To associate a localization length with $\lambda_i$, we write
\begin{equation}
\lambda_i = e^{A_1 + i B_2}
\end{equation}
such that $\psi_j$ has an envelope that goes as $e^{j {\rm max} (A_1, A_2)}$. The overall procedure is summarized as follows:
\begin{enumerate}
\item We solve $E | \psi \rangle = H_{eff} | \psi \rangle$ for $E$. 
\item We solve Eqn.~\ref{eqn:QHN-quad} for $\lambda_i(E)$. 
\item We now have $\lambda_i(E)$ for $i = 1, 2$. $A_i$ can be extracted from $| \lambda_i(E) | = e^{A_i}$.
\item We numerically diagonalize $H_{\rm HN}$ to obtain $E$, and use $E$ to obtain $\lambda_i$.
\end{enumerate}
We numerically confirm that the above procedure reproduces the localization length for the nearest-neighbor Hatano-Nelson model.

To apply the same method to the NNN Hatano-Nelson model, we write down the analogous eigenvalue equation
which has eigenvalue equation 
\begin{widetext}
\begin{equation}
(E + i ( \Gamma + \kappa ) ) \psi_j = \frac{w + \kappa}{2} \psi_{j - 1} + \frac{w - \kappa}{2} \psi_{j + 1} + \frac{T}{2} e^{i \phi} \psi_{j - 2} + \frac{T}{2} e^{-i \phi} \psi_{j + 2}
\end{equation}
\end{widetext}

Using the ansatz $\psi_n = c \lambda^n$, we have the equation 
\begin{equation}
\begin{aligned}
    0 &= \frac{T}{2} e^{-i \phi} \lambda^4+ \frac{w - \kappa}{2} \lambda^3 \\
    &\qquad - (E + i ( \Gamma + \kappa ) ) \lambda^2  + \frac{w + \kappa}{2} \lambda^1  + \frac{T}{2} e^{i \phi} 
\end{aligned}
\end{equation}
which has four independent solutions, so we write 
\begin{equation}
\psi_n = \sum_{i = 0}^3 c_i \lambda_i^n
\end{equation}

Following the same procedure as before we can obtain $\lambda_i(E)$, and associate each $\lambda_i$ with a complex exponent $A_i(E) + i B_i(E)$, and identify the shortest localization length with the largest exponent ${\rm max} (A_0, A_1, A_2, A_3)$.

\begin{figure}
    \centering
    \includegraphics[width=0.99\linewidth]{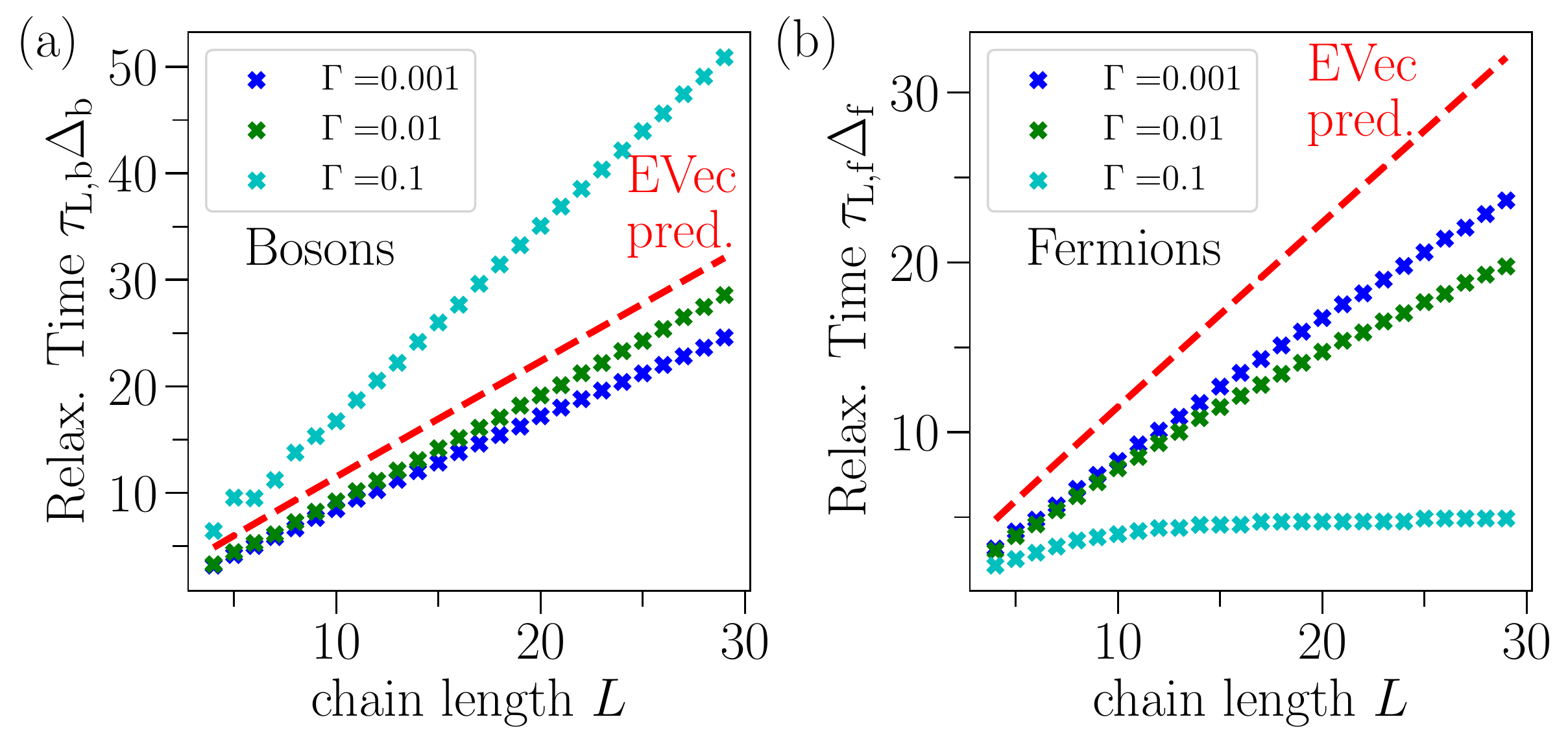}
    \caption{
    (a) The longest decay time $\tau_{\rm L, b}$ for bosons on a NNN QHN chain of length $L$ in units of dissipative gap $\Delta_{\rm b}$, with parameters given by $w = T = 1$, $\kappa = 0.999 w$ and $\phi = \pi / 2$ for various values of $\Gamma$. Numerically obtained values are given by the scatter plots whereas the EVec prediction is given by the dashed lines. Since $\Gamma$ does not affect the localization length, the EVec prediction collapses into the same line for all three values of $\Gamma$. In this case, the EVec prediction is a gets the linear in $L$ scaling right and appears to give a good order of magnitude estimate for the relaxation times. However, it does not account for the effects of pumping $\Gamma$. (b) Same as (a) but for fermions. We observe that for larger values of $\Gamma$, the relaxation times saturate, as with the usual QHN chain.}
    \label{fig:nr_NNN_combined_plots}
\end{figure}

In this manner, we plot the numerically simulated longest decay times and the EVec prediction in Fig.\ref{fig:nr_NNN_combined_plots} for a highly non-reciprocal chain with chosen parameters $\kappa = 0.999w$, $T = w$ and $\phi = \pi / 2$. In this example, the EVec scaling appears to coincide with the small $\Gamma$ behavior. However, it does not explain the $\Gamma$ dependence of the relaxation times, and fails dramatically for large $\Gamma$ in the fermionic case.

\bibliography{apssamp}

\end{document}